\documentstyle[12pt,fullpage]{article}

\newcommand{\be}{\begin{equation}}
\newcommand{\ee}{\end{equation}}
\newcommand{\ba}{\begin{eqnarray}}
\newcommand{\ea}{\end{eqnarray}}
\newcommand{\non}{\nonumber\\ }
\newcommand{\oa}{\bar{a}}
\newcommand{\ob}{\bar{b}}

\newcommand{\umu}{\underline{\mu}}
\newcommand{\unu}{\underline{\nu}}
\newcommand{\urho}{\underline{\rho}}
\newcommand{\usig}{\underline{\sigma}}
\newcommand{\ukap}{\underline{\kappa}}
\newcommand{\ulam}{\underline{\lambda}}
\newcommand{\ua}{\underline{a}}
\newcommand{\ub}{\underline{b}}
\newcommand{\uc}{\underline{c}}
\newcommand{\ud}{\underline{d}}
\newcommand{\Tr}{{\mbox{Tr}}}

\begin{document}

\renewcommand{\thefootnote}{\fnsymbol{footnote}}
\font\csc=cmcsc10 scaled\magstep1
{\baselineskip=14pt
 \rightline{
 \vbox{\hbox{UT-783}
       \hbox{YITP-97-37}
       \hbox{July 1997}
}}}

\vfill
\begin{center}
{\large\bf
Matrix Regularization of Open Supermembrane}\\
---towards M-theory five-brane via open supermembrane ---

\vfill

{\csc Kiyoshi EZAWA}\footnote{JSPS fellow}$^{1)}$\footnote{
      e-mail address : ezawa@yukawa.kyoto-u.ac.jp},
{\csc Yutaka MATSUO}$^{2)}$\footnote{
      e-mail address : matsuo@hep-th.phys.s.u-tokyo.ac.jp},
{\csc Koichi MURAKAMI}$^{1)}$\footnote{
      e-mail address : murakami@yukawa.kyoto-u.ac.jp}\\

\vskip.2in

$^{1)}${\baselineskip=15pt
\it  Yukawa Institute for Theoretical Physics \\
  Kyoto University, Sakyo-ku, Kyoto 606-01, Japan \\
}

\vskip.2in

$^{2)}${\baselineskip=15pt
\it Department of Physics, Faculty of Science, University of Tokyo,\\
     Hongo, Bunkyo-ku, Tokyo 113, Japan\\
}
\end{center}

\vfill
\setcounter{footnote}{0}
\renewcommand{\thefootnote}{\arabic{footnote}}

\begin{abstract}
We study open supermembranes in 11 dimensional
rigid superspace with 6 dimensional topological defects
(M-theory five-branes).
After rederiving in the Green-Schwarz formalism
the boundary conditions for open
superstrings in the type IIA theory,
we determine the boundary conditions for open
supermembranes by imposing
kappa symmetry and invariance under a fraction
of 11 dimensional supersymmetry. The result seems to imply the
self-duality of the three-form field strength on the five-brane
world volume.
We show that the light-cone gauge formulation
is regularized by a dimensional reduction of
a 6 dimensional N=1 super Yang-Mills theory with the
gauge group SO(N$\rightarrow\infty$). We also
analyze the SUSY algebra and BPS states in the light-cone gauge.

\end{abstract}

hep-th/9707200
\setcounter{footnote}{0}
\renewcommand{\thefootnote}{\arabic{footnote}}
\newpage
\vfill

\section{Introduction}

During the recent progress in the understanding of
nonperturbative properties of superstring theory,
M-theory has played an indispensable role.
While its formulation is still controversial,
it is widely believed to possess the following properties:
\begin{enumerate}
 \item the effective low energy theory is described by 11 dimensional
supergravity\cite{r:Wit}; 
 \item the compactification
on $S^{1}$ coincides with the  type IIA superstring theory\cite{r:Wit};
 \item the compactification on $S^{1}/{\bf Z}_{2}$
yields the $E_{8}\times E_{8}$ heterotic string theory\cite{r:HW}; 
 \item membranes and five-branes play central roles \cite{r:Wit5}.
\end{enumerate}

Let us focus on the five-branes.
While covariant formulations of M-theory five-branes
have been found recently \cite{r:APPS} \cite{r:HSW},
it is still important to look for their alternative description.
In the case of superstring theory, solitonic objects
carrying Ramond-Ramond charges are described as
Dirichlet branes on which open strings can end\cite{r:Pol}.
This leads us to the idea of describing M-theory five-branes
as ``Dirichlet-branes'' on which open supermembranes can end
\cite{r:Tow}. 
Becker and Becker \cite{r:BB} identified boundary conditions
for an open supermembrane in a particular kind of light-cone gauge. 

In this paper we make a further step towards
the description of five-branes by means of
open supermembranes.
In sec.II we investigate open supermembranes
in the 11 dimensional rigid superspace in the covariant
formalism\cite{r:BST}.
It is known that an open supermembrane cannot exist without
the topological defects on which the membrane can end
\cite{r:BST}\cite{r:BM}. We re-examine the analysis of ref.\cite{r:BM}
and show that the topological defects should be of dimension 2 (string),
6 (five-brane) or 10 (nine-brane). 
We determine the boundary conditions
for the open supermembrane in the presence of a three-form field
strength on the five-brane world volume by requiring the
invariance of the action under kappa-symmetry and 1/2 supersymmetry.
The result seems to imply that the three-form field strength
be self-dual. 
As an exercise before the analysis of open supermembranes, 
we investigate, in the covariant Green-Schwarz formalism\cite{r:GS},
open superstrings
in the type IIA theory in the presence of a field strength
on the D brane world volume.
We rederive the boundary conditions found in the 
light-cone gauge\cite{r:GG}.
In sec.III, we construct the light-cone gauge formalism of
the open supermembrane by mimicking the procedure for the
closed supermembrane\cite{r:dWHN}. The resulting theory
is a (0+1)-dimensional $N=8$ supersymmetric gauge theory 
with gauge group being that of area preserving diffeomorphisms
(APD) which preserve the boundary conditions.
We show that the theory is well-approximated by
a dimensional reduction to (0+1)-dimension of a
(5+1)-dimensional N=1 SO(N) super Yang-Mills theory
with a hypermultiplet in the rank-2 symmetric tensor
representation of SO(N). In this SO(N) regularization, however,
correspondence between surface integral and trace is not so
complete as in the U(N) regularization of
the closed membrane. Thus in subsec.III C we propose
an alternative regularization of the integration
based on the idea of a \lq\lq non-commutative cylinder''.
Sec.IV is devoted to the analysis of the
11D SUSY algebra and that of BPS states in the light-cone gauge.
In sec.V we discuss possible extensions of our results and some
remaining issues.

In Appendix A, the reader can find the convention and formulae
used in this paper.

\section{Boundary conditions for open supermembranes}

Boundary conditions for  open supermembranes were analyzed in
refs.\cite{r:BB}\cite{r:BM}
when the background is the 11 dimensional rigid superspace.\footnote{
Bosonic open membranes
were studied in ref.\cite{r:KY} and
a preliminary investigation of an open supermembrane
was made in ref.\cite{r:BST}.}
In this section we extend the analysis to the case
of a nonzero three-form
field strength on the world volume of a ``Dirichlet brane''
by using the covariant formalism of the supermembrane theory
\cite{r:BST}. We argue that
the field strength seems to be identified 
with that of the two-form
potential on the five-brane when the ``Dirichlet brane'' is 6 dimensional.
In order to get accustomed to its treatment,
we begin by examining boundary conditions for open superstrings
in the Green-Schwarz (GS) formalism of the type IIA theory
when there is a constant two-form field strength on
the D-brane world volume. 
The key to determine the boundary condition
resides in the kappa-symmetry and supersymmetry.

\subsection{Open superstrings in the type IIA theory}

Our starting point is the Green-Schwarz action of the type
IIA superstring\cite{r:GS}
\ba
S&=&\int_{\Sigma}d^{2}\sigma\sqrt{-g}+\int_{\Sigma}{\cal L}_{WZ}
+\int_{\partial\Sigma}A,\non
{\cal L}_{WZ}&=&-i\bar{\theta}\Gamma_{\mu}\Gamma_{11}d\theta
(dX^{\mu}-\frac{i}{2}\bar{\theta}\Gamma^{\mu}d\theta),\label{eq:IIA}
\ea
where $(X^{\mu}(\sigma),\theta^{\alpha}(\sigma))$
($\mu=0,1,\ldots,9;\alpha=1,2,
\ldots,32$) denotes the embedding of the string world sheet $\Sigma$
(with boundary $\partial\Sigma$) into the 10 dimensional type
IIA rigid superspace. We note that $\theta$ is a Majorana spinor
with $\bar{\theta}=\theta^{T}{\cal C}$.
$A=dX^{\mu}A_{\mu}+d\theta^{\alpha}A_{\alpha}$ is a one-form
on the world volume of the Dirichlet-brane.\footnote{
As for differential forms we use the convention of Wess-Bagger
\cite{r:WB}. In this paper we
do not make explicit distinction
between the differential form on the superspace (or on the
Dirichlet brane) and its pullback to the world sheet,
because which is used is usually obvious from the context.}
{}For simplicity we consider only the case that the one-form
is bosonic, i.e.,
$$
A_{\alpha}=0,\quad \partial_{\alpha}A_{\mu}=0.
$$
$g$ denotes the determinant of the induced metric on the
world sheet
\ba
g_{rs}&=&\Pi_{r}^{\mu}\Pi_{s}^{\nu}\eta_{\mu\nu}\quad(r,s=\tau,\sigma),
\non
\Pi^{\mu}_{r}&=&\partial_{r}X^{\mu}-i\bar{\theta}\Gamma^{\mu}
\partial_{r}\theta.
\ea

In the case of the closed superstring, the action (\ref{eq:IIA})
has symmetry under the 10 D type IIA super  Poincar\'{e} transformations,
world sheet reparametrizations and local fermionic transformations
(kappa-symmetry). Among them we only give the fermionic
transformations:
\ba
\delta_{\epsilon}X^{\mu}&=&-i\bar{\theta}\Gamma^{\mu}\epsilon,
\quad \delta_{\epsilon}\theta=\epsilon,\\
\delta_{\kappa}X^{\mu}&=&i\bar{\theta}\Gamma^{\mu}(1+\Gamma)\kappa,
\quad \delta_{\kappa}\theta=(1+\Gamma)\kappa,
\ea
where $\epsilon$ is a constant 10D Majorana spinor and
$\kappa$ is a 10D Majorana spinor which depends on the coordinates
of the world sheet. The matrix $\Gamma$ is defined as
\be
\Gamma=\frac{\epsilon^{rs}}{2\sqrt{-g}}\Pi^{\mu}_{r}\Pi^{\nu}_{s}
\Gamma_{\mu\nu}\Gamma_{11},
\ee
and is subject to
$$
(\Gamma)^{2}=I_{32}, \quad {\cal C}^{-1}\Gamma^{T}{\cal C}=\Gamma.
$$

The kappa-symmetry is particularly important
in the GS formalism because it, together with the world sheet
reparametrization invariance, guarantees the matching of
bosonic and fermionic degrees of freedom on the world sheet.
Thus, in order to preserve a fraction of supersymmetry,
we have to keep the kappa-symmetry even in the presence of
the world-sheet boundary $\partial\Sigma$.
In the following we will look for the boundary conditions
which preserve kappa-symmetry and 1/2 of space-time SUSY.

{}First we investigate the kappa-symmetry.
Because it is preserved in the absence of
$\partial\Sigma$, 
when we take the variation of the action under the local fermionic
transformation
we are left only with the boundary terms
\ba
\delta_{\kappa}S&=&\int_{\partial\Sigma}
\left[\frac{1}{2}(\bar{\theta}\Gamma_{\mu}\Gamma_{11}d\theta
\bar{\theta}\Gamma^{\mu}\delta_{\kappa}\theta+
\bar{\theta}\Gamma_{\mu}\Gamma_{11}\delta_{\kappa}\theta
\bar{\theta}\Gamma^{\mu}d\theta)
+i\bar{\theta}\Gamma_{\mu}\Gamma_{11}\delta_{\kappa}\theta
dX^{\mu} \right]\non
& &+\int_{\partial\Sigma}dX^{\mu}\delta_{\kappa}X^{\nu}F_{\nu\mu},
\ea
where $F_{\mu\nu}=2\partial_{[\mu}A_{\nu]}$ is the two-form field strength.
These boundary terms vanish if the following conditions
hold on $\partial\Sigma$:
\ba
dX^{\oa}&=&\bar{\theta}\Gamma^{\oa}d\theta=\bar{\theta}\Gamma^{\oa}
(1+\Gamma)\kappa=0,\non
\bar{\theta}\Gamma_{\umu}\Gamma_{11}(1+\Gamma)\kappa&=&
{}F_{\umu\unu}\bar{\theta}\Gamma^{\unu}(1+\Gamma)\kappa,\label{eq:kappaIIA}
\ea
where $\umu,\unu=0,1,\cdots,p$ and $\oa=p+1,\cdots,9$.

Next we see the conditions for unbroken SUSY.
As in the case of kappa-symmetry we find
\ba
\delta_{\epsilon}S&=&\int_{\partial\Sigma}
\left[-\frac{1}{6}(\bar{\theta}\Gamma_{\mu}\Gamma_{11}\epsilon
\bar{\theta}\Gamma^{\mu}d\theta+\bar{\theta}\Gamma^{\mu}\epsilon
\bar{\theta}\Gamma_{\mu}\Gamma_{11}d\theta)-i
\bar{\theta}\Gamma_{\mu}\Gamma_{11}\epsilon dX^{\mu}
\right]\non
& &+\int_{\partial\Sigma}dX^{\mu}(-i\bar{\theta}\Gamma^{\nu}\epsilon)
{}F_{\nu\mu}.
\ea
It vanishes if we set, on $\partial\Sigma$,
\ba
dX^{\oa}&=&\bar{\theta}\Gamma^{\oa}d\theta=\bar{\theta}\Gamma^{\oa}\epsilon
=0,\non
\bar{\theta}\Gamma_{\umu}\Gamma_{11}\epsilon
&=&F_{\umu\unu}\bar{\theta}\Gamma^{\unu}\epsilon. \label{eq:susyIIA}
\ea

We have derived eqs.(\ref{eq:kappaIIA}) and (\ref{eq:susyIIA})
{}from the kappa-symmetry and supersymmetry respectively.
These two equations are of the same 
{}form except that $(1+\Gamma)\kappa$ in the former is replaced with 
$\epsilon$ in the latter. Thus we find it natural to impose the
{}following boundary conditions on $\partial\Sigma$:
\ba
\delta X^{\oa}&=&\bar{\theta}\Gamma^{\oa}\delta\theta=0
\quad (\oa=p+1,\ldots,9),\non
\bar{\theta}\Gamma_{\umu}\Gamma_{11}\delta\theta
&=&F_{\umu\unu}\bar{\theta}\Gamma^{\unu}\delta\theta
\qquad (\umu=0,1,\cdots,p).\label{eq:DIIA1}
\ea
This represents the situation in which the open superstring
ends on a (p+1)-dimensional hyperplane,
namely, on a D p-brane.

In order that a fraction of space-time SUSY be unbroken,
however, it is necessary to rewrite the boundary conditions for
the fermion $\theta$ in a linear form.
This requirement allows only the even integer $p$.
In this case the above boundary conditions are rewritten as
\ba
X^{\oa}&=&\mbox{ const. },\non
\theta&=&e^{\frac{1}{2}Y_{\umu\unu}\Gamma^{\umu\unu}\Gamma_{11}}
(\Gamma_{11})^{\frac{p-2}{2}}\Gamma_{(p)}\theta \non
&=&e^{\frac{1}{4}Y_{\umu\unu}\Gamma^{\umu\unu}\Gamma_{11}}
(\Gamma_{11})^{\frac{p-2}{2}}\Gamma_{(p)}
e^{-\frac{1}{4}Y_{\umu\unu}\Gamma^{\umu\unu}\Gamma_{11}}\theta,
\label{eq:DIIA2}
\ea
where ${F_{\umu}}^{\unu}={\tanh(Y)_{\umu}}^{\unu}$ should be
constant so that (\ref{eq:DIIA2}) yields
(\ref{eq:DIIA1}). This result coincides with
that obtained in ref. \cite{r:GG}
(see also ref.\cite{r:BKOP}).

Boundary conditions for the remaining fields, namely $X^{\umu}$
and
$$
\theta^{(+)}\equiv\frac{1}{2}
(1+e^{\frac{1}{2}Y_{\umu\unu}\Gamma^{\umu\unu}\Gamma_{11}}
(\Gamma_{11})^{\frac{p-2}{2}}\Gamma_{(p))})\theta,
$$
are determined from the compatibility with equations of motion.
Namely, in deriving equations of motion from the variations
of the action, we should choose boundary conditions such that
boundary terms vanish.
This leads us to find
\be
\Phi^{\umu}\equiv
\sqrt{-g}g^{\sigma s}\Pi^{\umu}_{s}-{F^{\umu}}_{\unu}\Pi_{\tau}^{\unu}
=0.\label{eq:NeumannIIA}
\ee
In principle we can specify the boundary conditions
completely by exploiting (\ref{eq:NeumannIIA}), its compatibility
with kappa-symmetry: $\delta_{\kappa}\Phi^{\umu}=0$, and the
equation of motion for $\theta$:
\be
\Pi_{r}^{\mu}\Gamma_{\mu}(\sqrt{-g}g^{rs}\partial_{s}+\epsilon^{rs}
\Gamma_{11}\partial_{s})\theta=0.
\ee
In general, however, it is difficult to carry out this task
because the conditions are fairly 
non-linear.
We therefore restrict ourselves to the following two cases which are
relatively tractable.

(1)\underline{$F_{\umu\unu}=0$}. In this case we can reduce
the condition (\ref{eq:NeumannIIA})  to the
linear one
\be
\Pi_{\sigma}^{\umu}=0.
\ee
If we impose the remaining two conditions we can separate
the bosonic and the fermionic parts as
\be
\partial_{\sigma}X^{\umu}=\partial_{\sigma}\theta^{(+)}=0\quad\mbox{ on }
\partial\Sigma.
\ee

(2)\underline{Light-cone conformal  gauge}.
In this gauge
$$
X^{+}=\tau, \quad \Gamma^{+}\theta=0,\quad g_{rs}\propto\delta_{rs},
$$
eq.(\ref{eq:NeumannIIA}) is simplified as
\ba
\partial_{\sigma}X^{\ua}=F_{\ua\ub}\partial_{\tau}X^{\ub}+F_{\ua +},
\ea
where $X^{\pm}=\frac{1}{\sqrt{2}}(X^{1}\pm X^{0})$, and
$\ua=2,\cdots,p$. In order for eq.(\ref{eq:NeumannIIA}) to be compatible
with the light-cone gauge, we must have $F^{+\umu}=F_{-\umu}=0$.
In the light-cone gauge, the kappa-symmetry is gauge-fixed and
the equation of motion is simplified as $\partial_{\tau}\theta=
\Gamma_{11}\partial_{\sigma}\theta$. Thus we find
\be
[1+e^{\frac{1}{2}Y_{\ua\ub}\Gamma^{\ua\ub}\Gamma_{11}}
(\Gamma_{11})^{\frac{p-2}{2}}\Gamma_{(p)}]\partial_{\sigma}\theta=0
\quad\mbox{ on }\partial\Sigma.
\ee
We note that the compatibility with supersymmetry further
requires $F_{+\ua}$ to vanish.

\subsection{Open supermembranes}

Let us now investigate open supermembranes. We consider the case
in which a two-form gauge field
$B=\frac{1}{2}dX^{\mu}dX^{\nu}B_{\nu\mu}$
couples to the boundary of the membrane world volume.\footnote{
In general we can consider a coupling $S_{int}=-\int_{\Sigma}C$ of
the membrane world volume
to the three-form potential $C$ which is a member of the
11 dimensional supergravity multiplet. The two-form $B^{(0)}$ is introduced
in order to maintain the gauge invariance of the theory\cite{r:BM}.
Actually $S+S_{int}$ (with $B$ in (\ref{eq:GSaction})
replaced with $B^{(0)}$) is invariant under the gauge transformations
$\delta C=d\Lambda$ and $\delta B^{(0)}=\Lambda$, where $\Lambda$ is
a space-time two-form field.

Because we are now working in the 11
dimensional rigid superspace in which $dC=0$ (see e.g. \cite{r:BST}),
we can express the three-form by a pure gauge, $C=d\Lambda^{(0)}$,
and thus we can absorb it into the
two-form $B^{(0)}$.

{}From this consideration, we see that the two-form field $B$
in eq.(\ref{eq:GSaction}) should actually be regarded as the
gauge-invariant object $B^{(0)}-\Lambda^{(0)}$ and that the
three-form field strength $H=dB$ equals $dB^{(0)}-C$
which coincides with the gauge invariant field strength
introduced in\cite{r:Tow}.}
The relevant action is
\ba
S&=&-\int_{\Sigma}d^{3}\xi\sqrt{-g}+\int_{\Sigma}{\cal L}_{WZ}
+\int_{\partial\Sigma}B, \non
{\cal L}_{WZ}&=&\frac{i}{2}\bar{\theta}\Gamma_{\mu\nu}d\theta
\left[(dX^{\mu}-i\bar{\theta}\Gamma^{\mu}d\theta)dX^{\nu}
{}-\frac{1}{3}\bar{\theta}\Gamma^{\mu}d\theta\bar{\theta}\Gamma^{\nu}
d\theta\right],\label{eq:GSaction}
\ea
where $(X^{\mu}(\xi),\theta^{\alpha}(\xi))$ ($\mu=0,1,\ldots,10$; $\alpha
=1,2,\ldots,32$) denotes the embedding of the membrane world volume
$\Sigma$ (with boundary $\partial\Sigma$)\footnote{
We use the world-volume coordinate
$(\xi^{i})=(\tau,\sigma^{1},\sigma^{2})$,
among which $(\tau,\sigma^{2})$ and $\sigma^{1}$ are,
respectively, tangent and normal to the boundary $\partial\Sigma$.
As for the volume form we use the convention $d\xi^{i}d\xi^{j}d\xi^{k}
=\epsilon^{ijk}d^{3}\xi$.}
into the 11 dimensional
rigid superspace. As in the case of type IIA strings $\theta$
is a Majorana spinor with $\bar{\theta}=\theta^{T}{\cal C}$.
We mean by $g$ the determinant of the induced metric
\ba
g_{ij}&=&\Pi_{i}^{\mu}\Pi_{j}^{\nu}\eta_{\mu\nu},\quad(i,j=0,1,2)\non
\Pi_{i}^{\mu}&=&\partial_{i}X^{\mu}-i\bar{\theta}\Gamma^{\mu}
\partial_{i}\theta.
\ea 

In the case of a closed supermembrane,
the action (\ref{eq:GSaction}) is invariant under 11D super Poincar\'{e}
transformations, world volume reparametrizations, and local
fermionic transformations (kappa-symmetry).
We only give expressions for fermionic transformations
\ba
\delta_{\epsilon}X^{\mu}&=&-i\bar{\theta}\Gamma^{\mu}\epsilon,
\quad \delta_{\epsilon}\theta=\epsilon, \non
\delta_{\kappa}X^{\mu}&=&i\bar{\theta}\Gamma^{\mu}(1+\Gamma)\kappa,
\quad \delta_{\kappa}\theta=(1+\Gamma)\kappa,\label{eq:kappa}
\ea
where $\epsilon$ is a constant 11D Majorana spinor and
$\kappa$ is a 11D Majorana spinor which depends on $(\xi^{i})$.
The matrix $\Gamma$ is defined as
\be
\Gamma=\frac{\epsilon^{ijk}}{3!\sqrt{-g}}\Pi^{\mu}_{i}\Pi^{\nu}_{j}
\Pi^{\rho}_{k}\Gamma_{\mu\nu\rho},
\ee
and has the following properties
\be
(\Gamma)^{2}=I_{32},
\quad {\cal C}^{-1}\Gamma^{T}{\cal C}=\Gamma,
\quad \Pi_{i}^{\mu}\Gamma\Gamma_{\mu}=
\Pi^{\mu}_{i}\Gamma_{\mu}\Gamma=\frac{g_{im}}{2\sqrt{-g}}
\epsilon^{mkl}\Pi^{\nu}_{k}\Pi^{\rho}_{l}\Gamma_{\nu\rho}.
\ee

As in the string case kappa-symmetry is indispensable
if we want to keep a part of world volume supersymmetry.
In the following we determine the boundary conditions
by imposing the invariance under kappa-symmetry and
under a fraction of 11D SUSY.
Variations of the action (\ref{eq:GSaction})
under the transformations (\ref{eq:kappa})
are computed as
\ba
\delta_{\kappa}S&=&\int_{\partial\Sigma}\left[
\frac{i}{2}\bar{\theta}\Gamma_{\mu\nu}d\theta
(idX^{\mu}\bar{\theta}\Gamma^{\nu}\delta_{\kappa}\theta
+\frac{1}{3}\bar{\theta}\Gamma^{\mu}d\theta\bar{\theta}\Gamma^{\nu}
\delta_{\kappa}\theta)\right.\non
& &\left.+\frac{i}{2}\bar{\theta}\Gamma_{\mu\nu}\delta_{\kappa}\theta
(dX^{\mu}dX^{\nu}-i\bar{\theta}\Gamma^{\mu}d\theta dX^{\nu}
{}-\frac{1}{3}\bar{\theta}\Gamma^{\mu}d\theta\bar{\theta}\Gamma^{\nu}d\theta)
\right]\non
& &+\int_{\partial\Sigma}(-\frac{i}{2}dX^{\mu}dX^{\nu}H_{\mu\nu\rho}
\bar{\theta}\Gamma^{\rho}\delta_{\kappa}\theta),
\non
\delta_{\epsilon}S&=&\int_{\partial\Sigma}\left[
{}-\frac{i}{2}\bar{\theta}\Gamma_{\mu\nu}\epsilon(dX^{\mu}dX^{\nu}
{}-\frac{i}{3}\bar{\theta}\Gamma^{\mu}d\theta dX^{\nu}
{}-\frac{1}{15}\bar{\theta}\Gamma^{\mu}d\theta
\bar{\theta}\Gamma^{\nu}d\theta) \right. \non
& &\left.+\frac{1}{6}\bar{\theta}\Gamma^{\nu}\epsilon
\bar{\theta}\Gamma_{\mu\nu}d\theta(dX^{\mu}-\frac{i}{5}
\bar{\theta}\Gamma^{\mu}d\theta)\right]\non
& &+\int_{\partial\Sigma}\frac{i}{2}dX^{\mu}dX^{\nu}H_{\mu\nu\rho}
\bar{\theta}\Gamma^{\rho}\epsilon,
\ea
where $H=dB$ is the three-form field strength.
In order for $\delta_{\kappa}S$ and $\delta_{\epsilon}S$ to vanish
it is sufficient to set up the following boundary conditions on
$\partial\Sigma$:
\ba
\delta X^{\oa}&=&\bar{\theta}\Gamma^{\oa}\delta\theta=0,
\non
\bar{\theta}\Gamma_{\umu\unu}\delta\theta&=&
H_{\umu\unu\urho}\bar{\theta}\Gamma^{\urho}\delta\theta,
\label{eq:Dirichlet}
\ea
where $\umu=0,1,\ldots,p$ and $\oa=p+1,\ldots,10$.
The first equation represents the situation in which the open supermembrane
ends on a (p+1)-dimensional hyperplane-like topological
defect in the rigid 11D superspace.
However, this is not the whole story.
In order to keep a fraction of 11D supersymmetry,
the boundary conditions for $\theta$ have to be
rewritten in a linear form.
{}From the upper equation of (\ref{eq:Dirichlet}) we can infer
a natural candidate for the desired linear boundary condition
\ba
\theta&=&F(\Gamma^{\umu};H_{\umu\unu\urho})\Gamma_{(p)}\theta,
\quad F(\Gamma^{\umu};0)=I_{32}, \non
I_{32}&=&(F(\Gamma^{\umu};H_{\umu\unu\urho})\Gamma_{(p)})^{2}.
\ea
The third equation follows from the consistency of the first equation.
Note that $F(\Gamma^{\umu};H_{\umu\unu\urho})$ must be real
(in the Majorana representation)
because $\theta$ is a Majorana spinor.

We first consider the $H_{\umu\unu\urho}=0$ case.
{}From the consistency condition $(\Gamma_{(p)})^{2}=I_{32}$,
we find that $\frac{p(p+1)}{2}$ be odd. Moreover, in order to
reproduce eq.(\ref{eq:Dirichlet}), we have to set $p$ to be odd.
It implies that this theory admits only the
$(p+1)$ dimensional topological defects with
\be
p=1,5,9.
\ee
The $p=5$ case represents the M-theory five-brane
and $p=9$ is related to Ho\v rava-Witten's ``end-of-the-world 9-branes''
\cite{r:HW}. What is puzzling is the $p=1$ case.
It would be interesting to pursue it further.
In this paper, however, we mainly concentrate on the $p=5$ case.

Let us next consider the case of nonzero $H_{\mu\nu\rho}$.
Unlike in the string case the last condition of (\ref{eq:Dirichlet})
cannot be interpreted as rotation in the five-brane world volume,
and thus it is difficult to find out $F(\Gamma^{\umu};H_{\umu\unu\urho})$
which reproduces (\ref{eq:Dirichlet}).
In a special case in which $H_{\umu\unu\urho}$ satisfies
a ``self-duality'' condition, however, we can construct such $F$.
The boundary condition in this case turns out to be
\be
\theta=\exp\left(\frac{-1}{3}h_{\umu\unu\urho}\Gamma^{\umu\unu\urho}\right)
\Gamma_{(5)}\theta,\quad h_{\umu\unu\urho}=\frac{1}{3!}
\epsilon_{\umu\unu\urho\usig\ukap\ulam}h^{\usig\ukap\ulam}(=\mbox{const.}).
\label{eq:linearD}
\ee
After lengthy and tedious calculation which is outlined in
Appendix B, we find that eq.(\ref{eq:linearD}) reproduces
the last condition of eq.(\ref{eq:Dirichlet}) provided that
\be
H_{\umu\unu\urho}=4h_{\umu\unu\usig}{(1-2k)^{-1}}_{\urho}^{\usig}\, ,
\label{eq:Hh}
\ee
where $k^{\umu}_{\unu}=h^{\umu\urho\usig}h_{\unu\urho\usig}$.
The pattern of the breakdown of 11D SUSY
following from (\ref{eq:linearD}) agrees with
that obtained from the analysis of the
five-brane dynamics\cite{r:HSW}.\footnote{
Our $H_{\umu\unu\urho}$ seems to correspond to
$4{e_{a}}^{m}{e_{b}}^{n}{e_{c}}^{p}H_{mnp}$ in ref.\cite{r:HSW}.
Actually eq.(\ref{eq:Hh}) coincides with eq.(53) of ref.\cite{r:HSW}
if we take account of this correspondence and the fact that
$H=dB$ is rewritten as $dB^{(0)}-C$. Here $B^{(0)}$ is the
``bare'' two-form field on the five-brane and
$C$ is the three-form potential in 11 dimensions (see footnote 3). }

We remark that the derivation of the above result depends
heavily on the self-duality of $h_{\umu\unu\urho}$.
While we have not yet been able to provide a complete proof,
we strongly believe that eq.(\ref{eq:linearD}) is the unique possibility
of the linear boundary condition and that the ``self-duality''
of $H_{\mu\nu\rho}$ naturally follows from the requirement of
kappa-symmetry and space-time SUSY.

The remaining boundary conditions are determined by
investigating boundary terms arising from the action principle.
After some computation we find the following boundary condition
\be
\Phi^{\umu}\equiv\sqrt{-g}g^{1k}\Pi_{k}^{\umu}-{H^{\umu}}_{\unu\urho}
\Pi^{\unu}_{2}\Pi^{\urho}_{\tau}=0.
\ee
In principle we can completely determine the boundary
conditions by further imposing $\delta_{\kappa}\Phi^{\umu}=0$
and by considering the equation of motion:
$\Pi^{\mu}_{i}\Gamma_{\mu}(1-\Gamma)g^{ij}\partial_{j}\theta=0$.
In the case of nonzero $H_{\umu\unu\urho}$, however,
the conditions become highly nonlinear and it is difficult
to reduce them to a tractable form.
{}From now on we therefore consider
the case $H_{\umu\unu\urho}=0$ only.
In this case we can set the Neumann boundary conditions 
\ba
\partial_{1}X^{\umu}&=&0,\non
(1+\Gamma_{(p)})\partial_{1}\theta&=&0.\label{eq:Neumann}
\ea

Before concluding this section we consider the restriction on
the world volume reparametrization.
Under the infinitesimal reparametrization
$\xi^{i}\rightarrow\xi^{i}+v^{i}$, world volume fields transform as
$$
\delta_{v}X^{\mu}=v^{i}\partial_{i}X^{\mu},\quad
\delta_{v}(\partial_{i}X^{\mu})=v^{j}\partial_{j}(\partial_{i}X^{\mu})
+\partial_{i}v^{j}\partial_{j}X^{\mu},
$$
and the resulting variation of the action is
$$
\delta_{v}S=\int_{\Sigma}d^{3}\xi\partial_{i}(v^{i}{\cal L})
=-\int_{\partial\Sigma}d\tau d\sigma^{2}v^{1}{\cal L}.
$$
Imposing the conditions that $\delta_{v}S$ vanish and that
the boundary conditions (\ref{eq:Dirichlet}) (\ref{eq:Neumann})
be preserved by the above transformation, we find the following
boundary conditions for the generator $v^{i}$:
\ba
v^{1}\qquad&=&0,\non
\partial_{1}v^{0}=\partial_{1}v^{2}&=&0.
\ea

\section{Matrix regularization of an open supermembrane}

The matrix regularization of a closed supermembrane
was proposed by de Wit, Hoppe and Nicolai (dWHN)\cite{r:dWHN}.
{}Following their prescription we construct
the matrix regularization of an open supermembrane.
{}For simplicity, we investigate the case in which
there exist(s) either one or two parallel five-brane(s). 
In this situation, only DD and NN sectors appear;
therefore we need not consider either DN or ND sector.

\subsection{Light-cone gauge formulation}

Because the matrix regularization of the dWHN closed supermembrane
is based on the light-cone gauge formulation, 
we apply this formulation to the open supermembrane.
We will henceforth
use the notation $(\mu)=(+,-,a)$, $X^{\pm}=\frac{1}{\sqrt{2}}(X^{1}
\pm X^{0})$, $a=2,3,\ldots,10$,
and $(\xi^{i})=(\tau,\sigma^{r})$ ($r=1,2$).
The light-cone gauge is characterized by the conditions
\ba
X^{+}&=&\tau,\non
\Gamma^{+}\theta&=&0.\label{eq:light-cone}
\ea
{}Following dWHN we further impose the
conformal-like gauge conditions
\ba
g_{\tau r}&=&\partial_{r}X^{-}-i\bar{\theta}\Gamma^{-}\partial_{r}\theta
+\partial_{\tau}X^{a}\partial_{r}X^{a}=0,\non
g_{\tau\tau}&=&2(\partial_{\tau}X^{-}-i\bar{\theta}\Gamma^{-}
\partial_{\tau}\theta)+\partial_{\tau}X^{a}\partial_{\tau}X^{a}\non
&=&-\frac{1}{(P_{0}^{+}\sqrt{w})^{2}}\det(g_{rs})
=-\frac{1}{2(P_{0})^{2}}(\{X^{a},X^{b}\})^{2},
\label{eq:conformal}
\ea
where $\sqrt{w}(\sigma)$ is some fixed scalar density
on the constant-$\tau$ surface $\Sigma^{(2)}$ which is normalized
as $\int_{\Sigma^{(2)}}d^{2}\sigma\sqrt{w(\sigma)}=1$, $P^{+}_{0}$ is
a nonzero constant, and
\be
\{A,B\}=\frac{\epsilon^{rs}}{\sqrt{w}}\partial_{r}A\partial_{s}B
\ee
stands for the Lie bracket which generates area preserving
diffeomorphisms (APD) on $\Sigma^{(2)}$.
Substituting these gauge-fixing conditions into the
action (\ref{eq:GSaction}), we find
\be
S=\int d\tau\int_{\Sigma^{(2)}}
d^{2}\sigma\sqrt{w}\left[\frac{1}{2}P^{+}_{0}(\partial_{\tau}X^{a})^{2}
{}-iP^{+}_{0}\bar{\theta}\Gamma^{-}\partial_{\tau}\theta
{}-\frac{1}{4P^{+}_{0}}(\{X^{a},X^{b}\})^{2}+i\bar{\theta}\Gamma^{-}\Gamma^{a}
\{X^{a},\theta\}\right].
\ee
By using the nine dimensional spinor notation (see Appendix A.2),
it is rewritten as
\be
S=\int d\tau\int_{\Sigma^{2}}d^{2}\sigma\sqrt{w}
\left[\frac{P^{+}_{0}}{2}(\partial_{\tau}X^{a})^{2}+\frac{i}{2}\theta^{T}
\partial_{\tau}\theta-\frac{1}{4P^{+}_{0}}(\{X^{a},X^{b}\})^{2}
+\frac{i}{2P^{+}_{0}}\theta^{T}\gamma^{a}\{X^{a},\theta\}
\right],
\label{eq:LCaction}
\ee
where $\theta=(\theta^{\alpha})^{T}$ ($\alpha=1,2,\ldots.16$)
is now regarded as a real SO(9)  spinor.\footnote{
Relation between the SO(10,1) Majorana spinor and
the SO(9) real spinor is given by
$$
\theta|_{SO(10,1)}=\frac{1}{2^{3/4}\sqrt{P^{+}_{0}}}
\left(\begin{array}{c}0\\ \theta|_{SO(9)}\end{array}\right).
$$}

{}From the compatibility between the gauge-fixing conditions
(\ref{eq:light-cone})(\ref{eq:conformal})
and the boundary conditions obtained in the previous section:
\ba
\delta X^{\oa}&=&(1-\Gamma_{(5)})\theta=0,\non
\partial_{1}X^{\umu}&=&(1+\Gamma_{(5)})\partial_{1}\theta=0
\quad\mbox{ on }\partial\Sigma,
\ea
we see that $X^{\pm}$ are always parallel to the five-brane
world volume. Thus we introduce the notation $(\umu)=(+,-,\ua)$
with $\ua=2,3,4,5$.

In order to go further
we introduce a metric $w_{rs}(\sigma)$ on $\Sigma^{(2)}$
such that
\be
w(\sigma)=\det(w_{rs}(\sigma)),\quad
w_{12}|_{\partial\Sigma^{(2)}}=0,\quad
\partial_{r}(\sqrt{w}w^{r1})|_{\partial\Sigma^{(2)}}=0.
\ee
We can then perform the mode expansion
\ba
X^{\oa}(\sigma)&=&\sum_{A}Y_{A}^{(D)}(\sigma)X^{\oa A},\quad
\theta^{(+)}(\sigma)=\sum_{A}Y_{A}^{(D)}(\sigma)\theta^{(+)A},
\non
X^{\ua}(\sigma)&=&\sum_{A}Y_{A}^{(N)}(\sigma)X^{\ua A},\quad
\theta^{(-)}(\sigma)=\sum_{A}Y_{A}^{(N)}(\sigma)\theta^{(-)A},
\ea
where we have used the notation $\theta^{(\pm)}\equiv
\frac{1\pm\gamma_{(4)}}{2}\theta$ with $\gamma_{(4)}=\gamma^{2345}$.
We take $Y_{A}^{(D)}(\sigma)$ and $Y_{A}^{(N)}(\sigma)$ to be
eigenfunctions of the Laplacian
$$
\Delta Y^{(D,N)}_{A}\equiv
\frac{1}{\sqrt{w}}\partial_{r}\left(\sqrt{w}w^{rs}\partial_{s}
Y_{A}^{(D,N)}\right)=-\omega_{A}^{(D,N)}Y_{A}^{(D,N)}
$$
which are
subject to the Dirichlet and Neumann boundary conditions, respectively.
They can be chosen to satisfy the orthonormality
\ba
\int_{\Sigma^{(2)}}d^{2}\sigma\sqrt{w}Y_{A}^{(D)}(Y_{B}^{(D)})^{\ast}
&=&\delta_{A}^{B},\non
\int_{\Sigma^{(2)}}d^{2}\sigma\sqrt{w}Y_{A}^{(N)}(Y_{B}^{(N)})^{\ast}
&=&\delta_{A}^{B}.
\ea
In general, however, Dirichlet modes and Neumann modes are
not orthogonal to each other. While this guarantees the
completeness of $\{Y^{(D)}_{A}(\sigma)\}$ ($\{Y_{A}^{(N)}(\sigma)\}$)
in the space of functions on $\Sigma^{(2)}$ which
satisfy the Dirichlet (Neumann) boundary condition,\footnote{
{}For example, in the interval parametrized by
$\sigma^{1}\in[0,1/2]$,
$\cos(2\pi\sigma^{1})-\cos(6\pi\sigma^{1})$
can be expanded in terms of $\sin(2m\pi\sigma^{1})$.}
we need an extra care in order to discuss symmetry and dynamics
of the open supermembrane.

{}From the action (\ref{eq:LCaction}) we can derive the Dirac
brackets. We only pick the nonvanishing ones
\ba
(X^{\oa}(\sigma),P^{\ob}(\sigma^{\prime}))_{DB}&=&
\delta^{\oa\ob}\delta^{(D)}(\sigma,\sigma^{\prime}),\non
(\theta^{(+)}_{\alpha}(\sigma),\theta^{(+)}_{\beta}(\sigma^{\prime}))_{DB}
&=&\frac{-i}{\sqrt{w(\sigma)}}
\left(\frac{1+\gamma_{(4)}}{2}\right)_{\alpha\beta}
\delta^{(D)}(\sigma,\sigma^{\prime}),\non
(X^{\ua}(\sigma),P^{\ub}(\sigma^{\prime}))_{DB}&=&
\delta^{\ua\ub}\delta^{(N)}(\sigma,\sigma^{\prime}),\non
(\theta^{(-)}_{\alpha}(\sigma),\theta^{(-)}_{\beta}(\sigma^{\prime}))_{DB}
&=&\frac{-i}{\sqrt{w(\sigma)}}
\left(\frac{1-\gamma_{(4)}}{2}\right)_{\alpha\beta}
\delta^{(N)}(\sigma,\sigma^{\prime}),\label{eq:Dirac}
\ea
where we have defined $P^{a}\equiv P_{0}^{+}\sqrt{w}\partial_{\tau}
X^{a}$ and
$$
\delta^{(D,N)}(\sigma,\sigma^{\prime})\equiv\sqrt{w(\sigma)}
\sum_{A}Y_{A}^{(D,N)}(\sigma)
(Y_{A}^{(D,N)}(\sigma^{\prime}))^{\ast}.
$$
Time evolution of the system is described by the Hamiltonian
\ba
H&=&\int_{\Sigma{(2)}}d^{2}\sigma\frac{\sqrt{w}}{P_{0}^{+}}
\left[\frac{(P^{a})^{2}}{2w}+\frac{1}{4}(\{X^{a},X^{b}\})^{2}
{}-\frac{i}{2}\theta^{T}\gamma^{a}\{X^{a},\theta\}\right]\non
&=&\frac{(P_{0}^{\ua})^{2}+{\cal M}^{2}}{2P^{+}_{0}},
\label{eq:Hami}
\ea
where $P^{\ua}_{0}=\int_{\Sigma^{(2)}}d^{2}\sigma P^{\ua}(\sigma)$
is the total momentum along the five-brane world volume. We can also
regard this equation as the definition of
the invariant squared mass ${\cal M}^{2}$ of the open supermembrane.
We note that the first equation of (\ref{eq:conformal}) implies,
as integrability conditions of $X^{-}$,
\ba
\varphi(\sigma)&=&-\{\frac{P^{a}}{\sqrt{w}},X^{a}\}-\frac{i}{2}
\{\theta^{T},\theta\}\approx0,\label{eq:APDconstraint} \\
\varphi_{\lambda}&=&\int d^2\sigma {\phi^{(\lambda)}}^{r}\left(
P^{a}\partial_{r}X^{a}+\frac{i}{2}\sqrt{w}\theta^{T}\partial_{r}
\theta\right) \approx 0\, ,
\ea
where $\{\phi^{(\lambda)}_{r}\}$ is a basis of
$H^{1}(\Sigma^{(2)};{\bf R})$. 
They are regarded as first class constraints which
generates area preserving diffeomorphisms
\ba
\delta_{\zeta}X^{a}&=&\{\zeta,X^{a}\},\non
\delta_{(\lambda)}X^{a}&=&\phi^{(\lambda)r}\partial_{r}X^{a}.
\label{eq:APDtfm}
\ea
Due to the consideration in
the last of subsec.II B, the APD parameter $\zeta(\sigma)$ is
subject to the Dirichlet boundary condition
\be
\zeta(\sigma)=0 \quad\mbox{ on }\partial\Sigma.
\ee

As in the closed supermembrane case\cite{r:dWHN} the
theory we constructed 
can be regarded as a $(0+1)$-dimensional gauge theory
with gauge group being the group of area preserving diffeomorphisms
which preserve the boundary conditions.
The action of the gauge theory is
\be
S=\int dt\int_{\Sigma^{(2)}}d^{2}\sigma\sqrt{w}
\left[\frac{1}{2}(D_{t}X^{a})^{2}+\frac{i}{2}\theta^{T}D_{t}\theta
{}-\frac{1}{4}(\{X^{a},X^{b}\})^{2}+\frac{i}{2}\theta^{T}\gamma^{a}\{X^{a},
\theta\}\right],\label{eq:APDaction}
\ee  
where we have used the covariant derivative such as
$D_{t}X^{a}=\partial_{t}X^{a}-\{\omega,X^{a}\}$.
The gauge transformation of the APD connection $\omega(t,\sigma)$
is given by
\be
\delta_{\zeta}\omega=D_{t}\zeta=\partial_{t}\zeta-\{\omega,\zeta\},
\ee
where the gauge parameter $\zeta$ depends on time in general.
We note that, since the connection is a Lie algebra-valued 1-form,
it satisfies the Dirichlet boundary condition
\be
\omega(t,\sigma)=0\quad\mbox{ on }\partial\Sigma.
\ee

We see that the Hamiltonian (\ref{eq:Hami}) and the constraint
(\ref{eq:APDconstraint}) coincide with those derived from the action
(\ref{eq:APDaction}) {\em in the $\omega=0$ gauge}.\footnote{
Actually the action (\ref{eq:LCaction}) is equivalent to
(\ref{eq:APDaction}) in the $\omega=0$ gauge if we realize the time
rescaling $t=\tau/P^{+}_{0}$.
}
The residual gauge symmetry is 
the time-independent
APD transformations (\ref{eq:APDtfm}).

As we will see in sec.IV this theory has dynamical supersymmetry
generated by the supercharge $Q^{+}_{(+)}$ which has eight components.
Our theory is therefore regarded as a (0+1)-dimensional $N=8$
supersymmetric gauge theory.

\subsection{SO(N) regularization}

In ref.\cite{r:dWHN}, de Wit, Hoppe and Nicolai have shown
that the closed supermembrane theory is well
approximated by U(N) supersymmetric
gauge theory in (0+1)-dimension. In this subsection we show
that the open supermembrane theory is well approximated 
by SO(N) supersymmetric gauge theory in (0+1)-dimension.
While we only deal with a cylindrical membrane,
our analysis can be extended to a general
topology if we exploit the mirror image prescription.

{}First we examine how the matter contents are approximated
by SO(N). We introduce the coordinates
$\sigma^{1}\in[0,\frac{1}{2}]$
and $\sigma^{2}(\sim\sigma^{2}+1)$ which respectively parametrize
the $I$- and the $S^{1}$-directions of the cylinder $I\times S^{1}$.
It should be kept in mind that we use the following fiducial metric
$$
(w_{rs})=\left(\begin{array}{cc}4&0\\0&1\end{array}\right).
$$
Because we are dealing  with the case in which
the open supermembrane ends on one five-brane (or on two parallel ones),
we have two kinds of mode functions
\ba
\mbox{Dirichlet (DD sector)}&:&
Y_{A}^{(D)}(\sigma^{1}=0,\sigma^{2})=
Y_{A}^{(D)}(\sigma^{1}=1/2,\sigma^{2})=0,
\non
\mbox{Neumann (NN sector)}&:&
\partial_{1}Y_{A}^{(N)}(\sigma^{1}=0,\sigma^{2})=
\partial_{1}Y_{A}^{(N)}(\sigma^{1}=1/2,\sigma^{2})=0.
\ea
 
The Dirichlet and Neumann modes are given by
\ba
Y_{A}^{(D)}(\sigma)&=&\sqrt{2}e^{2\pi iA_{2}\sigma^{2}}
\sin(2\pi A_{1}\sigma^{1}),\non
Y_{A}^{(N)}(\sigma)&=&\sqrt{2}e^{2\pi iA_{2}\sigma^{2}}
\cos(2\pi A_{1}\sigma^{1})\quad\mbox{ for } A_{1}\neq0,\non
Y_{(0,A_{2})}^{(N)}(\sigma)&=&\qquad e^{2\pi iA_{2}\sigma^{2}}.
\ea
This leads us to find the following important correspondence (see Appendix C):
\footnote{Similar correspondence relation has been found by
Kim and Rey in a slightly different context\cite{r:KR}.}
\ba
\mbox{Dirichlet modes}&\stackrel{N\rightarrow\infty}{\longleftarrow}&
N\times N\mbox{ antisymmetric matrices},\non
\mbox{Neumann modes}&\stackrel{N\rightarrow\infty}{\longleftarrow}&
N\times N\mbox{ symmetric matrices}.\label{eq:correspondence}
\ea
Group structure of the APD gives another support for
the SO(N$\rightarrow\infty$) approximation. 
Namely, the Lie bracket has the following property
\ba
\{\mbox{Dirichlet},\mbox{Dirichlet}\}&\sim&\mbox{Dirichlet},\non
\{\mbox{Dirichlet},\mbox{Neumann}\}&\sim&\mbox{Neumann},\non
\{\mbox{Neumann},\mbox{Neumann}\}&\sim&\mbox{Dirichlet}.\label{eq:APD}
\ea
This coincides with the structure of commutation relations
for $N\times N$ matrices.
Actually we can see that the large $N$ limit of the commutation
relations reproduces the corresponding APD brackets.

However, we need a careful consideration with regard to the
matrix regularization of the action, constraints and
conserved charges. As is already pointed out in subsec.III A,
Dirichlet modes and Neumann ones are not orthogonal
w.r.t. the integration $\int_{\Sigma^{(2)}}d^{2}\sigma$,
while antisymmetric matrices and symmetric ones are
orthogonal to each other w.r.t. the trace of $N\times N$ matrices.
This tells us that we cannot naively replace the integral
$\int_{\Sigma^{(2)}}d^{2}\sigma\sqrt{w} A(\sigma)B(\sigma)$ by the trace 
${\rm Tr}(AB)$.
In the next subsection we will propose a few 
redefinitions of the ``trace'' which should be used
to define the Lagrangian.
However, the following argument shows that
we can nevertheless use the original ``naive'' definition of
the trace as long as we use it to approximate the integral in
the Lagrangian, smeared constraints, and conserved charges.

By inspecting the action (\ref{eq:LCaction}),
we find that it has the structure
\be
\int_{\Sigma^{(2)}}d^{2}\sigma\sqrt{w}
\left[(\mbox{Dirichlet})\times(\mbox{Dirichlet})
+(\mbox{Neumann})\times(\mbox{Neumann})\right],\label{eq:DDNN}
\ee
and that it never contains terms like $\int_{\Sigma^{(2)}}d^{2}\sigma
\sqrt{w}(\mbox{Dirichlet})\times(\mbox{Neumann})$.
This is  true also for constraints and conserved charges because 
of the following reasoning. They
become generators of some transformations and physically 
relevant transformations
must preserve the boundary conditions.
{}From the Dirac brackets (\ref{eq:Dirac}) we see that such generators
must be of the structure (\ref{eq:DDNN}).
What remains to be shown is that the integration
$\int_{\Sigma^{(2)}}d^{2}\sigma\sqrt{w}$ indeed has properties of the
trace for these restricted situations.
This is shown in Appendix D.

We can now give the explicit form of the
regularized theory. We replace the real functions on $\Sigma^{(2)}$
by $N\times N$ hermitian matrices.
The action and the constraint are given by
replacing $\int_{\Sigma^{(2)}}d^{2}\sigma\sqrt{w}(A(\sigma)B(\sigma))$
and $\{A,B\}$ in eqs.(\ref{eq:APDaction}) and (\ref{eq:APDconstraint})
with ${\rm Tr}(AB)$ and $i[A,B]$, respectively.
We find
\ba
S&=&\int dt
\mbox{Tr}\left(\frac{1}{2}(D_{t}X^{a})^{2}+\frac{i}{2}
\theta^{T}D_{t}\theta+\frac{1}{4}([X^{a},X^{b}])^{2}-\frac{1}{2}
\theta^{T}\gamma^{a}[X^{a},\theta]\right),
\non
\varphi&=&-i[P^{a},X^{a}]+\frac{1}{2}[\theta^{\alpha},\theta^{\alpha}]_{+},
\label{eq:matrix}
\ea
where 
we have defined
the covariant derivative $D_{t}A=\partial_{t}A-i[\omega,A]$
with an SO(N) connection $\omega(t)$, and introduced
the anticommutator $[A,B]_{+}\equiv AB+BA$ of the matrices $A$ and $B$.
$P^{a}=D_{t}X^{a}$
is the  momentum conjugate to $X^{a}$.
Needless to say, the Gauss law constraint $\varphi$ generates
SO(N) gauge transformations.
In terms of the SO(N) representation, the matter contents are
classified as
\ba
\mbox{Adjoint}\qquad&:&\omega,X^{6},X^{7},X^{8},X^{9},X^{10},\quad\theta^{(+)},
\non
\mbox{Symmetric rank-2}&:&\quad
X^{2},X^{3},X^{4},X^{5},\quad\qquad\theta^{(-)}.
\label{eq:matter}
\ea
We note that the fermion $\theta^{(+)}$ (or
$\theta^{(-)}$) corresponds to the 4 real canonical pairs
and that the two bosonic canonical pairs in the adjoint representation
are absorbed into gauge degrees of freedom. Thus we find
that, up to a finite number associated with zero-modes,
bosonic and fermionic degrees of freedom
precisely match with each other and thus  supersymmetry
is expected to hold in a rigorous sense.

We may interpret the matter content (\ref{eq:matter}) in terms of
a (5+1)-dimensional theory. The adjoint matter corresponds to
the 6D N=1 vector multiplet and thus
it is  considered to be obtained from a
6D N=1 super Yang-Mills field. The matter in the symmetric
representation is regarded as coming from a 6D N=1 hyper multiplet.
Let us  next consider the number of the generators of the dynamical
supersymmetry.
The dWHN closed supermembrane has 16 generators of dynamical supersymmetry,
corresponding to N=1 SUSY in 10D.\footnote{
This type of supersymmetric gauge quantum mechanics was first
discussed in ref.\cite{r:CH}
}
In the open supermembrane case, symmetry generated by
a half of them is broken due to the
boundary (see sec.IV). 
Therefore this theory has dynamical supersymmetry generated
by 8 supercharges, corresponding
to N=1 SUSY in 6D.
{}From these  indications it would be plausible 
to consider the matrix theory(\ref{eq:matrix})
to be the dimensional reduction to (0+1)-dimension of the
(5+1)-dimensional $SO(N)$ N=1 supersymmetric Yang-Mills theory with
a hyper multiplet in the rank-2 symmetric representation.
\footnote{When we quantize the fermionic zero-modes
  $\theta_{0}^{(-)} \sim {\rm Tr}\theta^{(-)}$,
  we have a 6D N=1 tensor multiplet and a hypermultiplet
  both of which are $SO(N)$ singlet \cite{r:BB}.
  They are expected to yield the collective coordinates of 
  the five-brane \cite{r:CHS}.}

To conclude this subsection we make a remark. In the cylindrical membrane,
there is one more constraint associated with rotation
along the $S^{1}$-direction. It plays an important role
if we compactify the dimensions
parallel to the five-brane world volume. It is given by
\be
\varphi_{2}=\int d^{2}\sigma\sqrt{w}\left[P^{+}_{0}\partial_{2}X^{-}
+\frac{P^{a}}{\sqrt{w}}\partial_{2}X^{a}
+\frac{i}{2}\theta^{T}\partial_{2}\theta\right]\approx 0.
\ee
As in ref.\cite{r:EMM} it is in principle possible to consider
a matrix version of this constraint.
In this paper, however, we will not pursue this issue any more.

\subsection{Regularization via a non-commutative cylinder}

In the last subsection we proposed an SO(N) regularization
of the open supermembrane. As we have seen, however,
we cannot obtain the complete correspondence between
the integration and the trace in this regularization.
This is unsatisfactory if one wants to regularize
the theory of open supermembranes by means of a
``non-commutative cylinder''.
In this subsection we give a few proposals to modify the
definition of the trace to make a
consistent correspondence with
the integral. The ambiguity comes from the formula,
\begin{equation}\label{e:quasiFourier}
\int_0^{1/2} d\sigma e^{2\pi i m \sigma} =
\left\{
\begin{array}{ll}
1/2\qquad&  m=0\\
i\frac{1-(-1)^m}{2\pi m} \qquad & m\neq 0.
\end{array}
\right.
\end{equation}
Obviously the usual definition of the trace gives
$Tr U^m = N\delta_{m0}$ (mod $N$) and it cannot give the second term
of (\ref{e:quasiFourier}).

To begin with we show that
the modified definition of the trace $\Tr'$ can not satisfy
the fundamental property of the trace,
\be\label{e:commutationInTrace}
\Tr'(AB) = \Tr'(BA).
\ee
This is because the commutation
relation $VUV^{-1}=\omega U$ ($\omega=e^{2\pi i /N}$) and
$\Tr' U\neq 0$ are not consistent with (\ref{e:commutationInTrace}).
In this sense, the problem is similar to the definition
of $p$ and $q$ with $\left[q,p\right]= I_N$
whose realization is impossible in the finite $N$.

This observation leads us to give a modified definition
of the trace as follows.  Let us consider the case when
$N$ is even, namely $N\equiv 2M$.  The modified definition of the
trace may be given as
\be
\Tr' A = \Tr {\cal P} A,
\quad
{\cal P} = \left(\begin{array}{cc}
I_M & 0_M\\ 0_M & 0_M \end{array}\right).
\ee
It gives
\be
\frac{1}{N}\Tr' U^m V^n= 
\left\{ 
\begin{array}{ll}
\frac{1}{2}\delta_{n0} \qquad & m=0 \quad\mbox{mod}\ N\\
\delta_{n0}\frac{1-(-1)^m}{N(1-\omega^m)} \qquad & m\neq 0 \quad
\mbox{mod}\ N
\end{array}
\right. .
\ee
This is obviously consistent with (\ref{e:quasiFourier})
in the large $N$ limit. Although the relation (\ref{e:commutationInTrace})
is violated in the finite $N$, the anomalous components appear only at
the ``boundary'' of the $M\times M$ blocks.
Such terms are supposed to disappear in the large $N$ limit.

Another possible redefinition is to keep the definition of the
trace but to redefine the generator which corresponds to
$e^{2\pi i\sigma_1/N}$.  Instead of using $U$, we introduce
the square root,
\be
U' = \left(
\begin{array}{cccc}
1 & 0 & \cdots & 0\\
0 & \omega' & \cdots & 0\\
\vdots & & \ddots & \vdots\\
0 & & \cdots & (\omega')^{N-1}
\end{array}
\right)
\qquad
\omega' = e^{\pi i /N}.
\ee
By easy manipulation,  one can prove the
consistency with (\ref{e:quasiFourier}) and 
(\ref{e:commutationInTrace}).  However, the commutation
relation $ VU' = \omega' U' V$ is violated.
As in the previous redefinition, the anomalous components 
appear at the boundary of the matrix and will disappear
in the large $N$ limit. 

The relation with the last subsection is clearer
in the first redefinition.  What we have proved in sec.III B is that,
in the definition of the lagrangian etc., one may effectively
replace $\Tr'$ with the ordinary trace since there are
no integrand of the (Dirichlet)$\times$(Neumann) type.

\section{11D SUSY algebra in the light-cone gauge}

In this section we investigate the supersymmetry
algebra of the model constructed in subsec.III A.
Extension to the matrix version is straightforward.

In the case of a closed supermembrane, there are two kinds of
supercharges
\ba
Q^{-}&=&\sqrt{P^{+}_{0}}\int d^{2}\sigma\sqrt{w}\theta, \non
Q^{+}&=&\frac{1}{\sqrt{P^{+}_{0}}}\int d^{2}\sigma
\left(P^{a}\gamma_{a}+\frac{\sqrt{w}}{2}\{X^{a},X^{b}\}\gamma_{ab}
\right)\theta.
\ea
They respectively generate kinematical SUSY transformations
\be
\delta_{-}X^{a}=0,\quad \delta_{-}\theta=\epsilon^{\prime},
\ee
and dynamical SUSY transformations
\be
\delta_{+}X^{a}=\epsilon\gamma^{a}\theta,\quad
\delta_{+}\theta=+i\left(\frac{P^{a}}{\sqrt{w}}\gamma_{a}
{}-\frac{1}{2}\{X^{a},X^{b}\}\gamma_{ab}\right)\epsilon,
\ee
where $\epsilon^{\prime}$ and $\epsilon$ are constant
real spinors of SO(9).

In the case of an open supermembrane, however, physically relevant
SUSY transformations have to preserve boundary conditions.
We are thus left with the following generators
\ba
Q^{-}_{(-)}&\equiv&\frac{1-\gamma_{(4)}}{2}Q^{-},\non
Q^{+}_{(+)}&\equiv&\frac{1+\gamma_{(4)}}{2}Q^{+}.
\label{eq:UBSUSY}
\ea
By using the Dirac brackets (\ref{eq:Dirac}) we have confirmed that
these generators are well-defined and that the action
(\ref{eq:LCaction}) is invariant under the SUSY transformations generated
by them even if the boundary terms are taken into account.
This is not surprising because these generators are of the form
(\ref{eq:DDNN}).

The algebra formed by the generators (\ref{eq:UBSUSY}) of
unbroken SUSY is found to be
\ba
i(Q^{-}_{(-)\alpha},Q^{-}_{(-)\beta})_{DB}&=&
({\cal P}^{(-)})_{\alpha\beta}P^{+}_{0},\non
i(Q^{-}_{(-)\alpha},Q^{+}_{(+)\beta})_{DB}&=&
({\cal P}^{(-)}\gamma_{\ua})_{\alpha\beta}P^{\ua}_{0}
+({\cal P}^{(-)}\gamma_{\oa\ub})_{\alpha\beta}
Z^{\oa\ub},\non
i(Q^{+}_{(+)\alpha},Q^{+}_{(+)\beta})_{DB}&=&
2H({\cal P}^{(+)})_{\alpha\beta}+2({\cal
P}^{+}\gamma_{\oa})_{\alpha\beta}
Z^{\oa},\label{eq:SUSYALG}
\ea
where we have introduced the projection operators ${\cal P}^{(\pm)}\equiv
\frac{I_{16}\pm\gamma_{(4)}}{2}$, and membrane charges are given by
\ba
Z^{ab}&=&-\int d^{2}\sigma\sqrt{w}\{X^{a},X^{b}\},\non
Z^{a}&=&\frac{1}{P^{+}_{0}}\int d^{2}\sigma\sqrt{w}\left(
\{X^{a},X^{b}\}\frac{P^{b}}{\sqrt{w}}+\frac{i}{2}\theta^{T}\{X^{a},\theta\}
\right)\non
&=&-\int d^{2}\sigma\sqrt{w}\{X^{a},X^{-}\}.
\ea
In the case of the closed supermembrane
we can use $Q^{+}$ and $Q^{-}$ to construct the
11 dimensional SUSY generator
\be
Q\equiv\left(\begin{array}{c}\sqrt{2}Q^{-}\\Q^{+}
\end{array}\right).
\ee
The generators of unbroken SUSY (\ref{eq:UBSUSY})
in the open membrane case
are then interpreted as those resulting from the projection
\be
\tilde{Q}\equiv\left(\begin{array}{c}\sqrt{2}Q^{-}_{(-)}\\
Q^{+}_{(+)}
\end{array}\right)
=\frac{1-\Gamma_{(5)}}{2}Q.
\ee
By virtue of this fact the SUSY algebra
(\ref{eq:SUSYALG}) is rewritten in a 11 dimensional form
\ba
&&i(\tilde{Q},\tilde{Q}^{T})_{DB}=
\left(\begin{array}{cc}
2P^{+}_{0}{\cal P}^{(-)}&\sqrt{2}{\cal P}^{(-)}(\not{P}+\not{Z}_{(2)})\\
\sqrt{2}{\cal P}^{(+)}(\not{P}-\not{Z}_{(2)})&2{\cal P}^{(+)}
(H\cdot I_{16}+\not{Z}_{(1)})
\end{array}\right)\label{eq:susyalg}\\
&&=\left(\begin{array}{cc}\sqrt{2}P^{+}_{0}{\cal P}^{(-)}&0\\
{\cal P}^{(+)}(\not{P}-\not{Z}_{(2)}) & {\cal P}^{(+)}
\end{array}\right)\left(\begin{array}{cc}
\frac{1}{P^{+}_{0}}{\cal P}^{(-)}&0\\ 0 & \frac{1}{P^{+}_{0}}{\bf m}
\end{array}\right)\left(\begin{array}{cc}
\sqrt{2}P^{+}_{0}{\cal P}^{(-)}&{\cal P}^{(-)}(\not{P}+\not{Z}_{(2)})\\
0&{\cal P}^{(+)}
\end{array}\right),\nonumber
\ea 
where we have defined the matrices
$\not{P}=P^{\ua}_{0}\gamma_{\ua}$, $\not{Z}_{(2)}=Z^{\oa\ub}
\gamma_{\oa\ub}$, $\not{Z}_{(1)}=Z^{\oa}\gamma_{\oa}$, and
\ba
{\bf m}&\equiv&{\cal P}^{(+)}\left[2P^{+}_{0}(H\cdot I_{16}+\not{Z}_{(1)})
{}-(\not{P}-\not{Z}_{(2)})(\not{P}+\not{Z}_{(2)})\right]\non
&=&{\cal P}^{(+)}\left[({\cal M}^{2}-Z^{\oa\ub}Z^{\oa\ub})I_{16}
+2(Z^{\oa}P^{+}_{0}+Z^{\oa\uc}P_{0}^{\uc})\gamma_{\oa}-Z^{\oa\ub}
Z^{\bar{c}\ud}
\gamma_{\oa\bar{c}\ub\ud}\right].\label{eq:determinant}
\ea

Here we make a comment on the five-brane charges.
In the present case, 11D SUSY generated by $Q$ is broken
to that generated by $\tilde{Q}=
\frac{1-\Gamma_{(5)}}{2}Q$,
as is expected from the result in subsec.II B.  
This pattern of breakdown of
supersymmetry coincides with that in the presence of
five-branes with charges
$$
P^{0}=Z^{12345}(=\infty).
$$
In this sense we can say that our theory of open supermembrane
effectively incorporates the longitudinal five-brane charge
$Z^{-2345}$ ($=Z^{+2345}$). We should remark that
we cannot incorporate the transverse five-brane charge
because the  gauge-fixing conditions
(\ref{eq:light-cone})(\ref{eq:conformal})
imply that $X^{\pm}$ be subject to the Neumann boundary condition.
This partially agrees with the statements in ref.\cite{r:BSS}.
There is, however, an essential difference. 
Namely we cannot give an explicit expression of the five-brane charge
{}from the bulk
such as $Z^{-abcd}\sim{\rm Tr}(X^{[a}X^{b}X^{c}X^{d]})$
which was discussed in \cite{r:BSS}. This is because the
corresponding expression in the membrane theory
$$
\int d^{2}\sigma\sqrt{w}\{X^{[a},X^{b}\}\{X^{c},X^{d]}\},
$$
always vanishes due to the identity $\epsilon^{r[s}\epsilon^{tu]}=0$.
We should rather identify the five brane charge
as coming from the topological defect at the boundary.

\subsection{BPS configurations}

Now that we have the 11D SUSY algebra, let us explore BPS
conditions. From the analysis of the SUSY algebra we see that
the nontrivial BPS configurations with nonvanishing membrane charges
should stretch both in the directions parallel and perpendicular
to the five-brane(s).
It is therefore sufficient to
consider the situation in which there are two parallel five-branes
and an open membrane which stretches between them.
Without loss of generality we can set the boundary conditions
\be
X^{\oa}|_{\sigma^{1}=0}=0,\quad
X^{\oa}|_{\sigma^{1}=1/2}=b\delta^{\oa}_{10},\quad
\partial_{1}X^{\ua}|_{\sigma^{1}=0}=
\partial_{1}X^{\ua}|_{\sigma^{1}=1/2}=0,
\ee
and so on.
In order to obtain nonvanishing membrane charges we further have to consider
either the case in which: (i) the membrane stretches infinitely
along the five-branes; or (ii)  the membrane wrapps
around a 1-dimensional cycle which is parallel to the five-branes.
We analyze the case (ii) in order to avoid the divergence of
the membrane charges.
Because we are dealing with flat five-branes we toroidally compactify
the directions parallel to the five-branes
\be
X^{\ua}\sim X^{\ua}+2\pi R^{\ua}, \quad X^{-}\sim X^{-}+2\pi R.
\ee
This also serves as a regularization of the case (i).

Thus we have $P^{+}_{0}=\frac{m}{R}$ and
\ba
X^{\ua}&=&\frac{Rm^{\ua}}{R^{\ua}m}\tau+2\pi R^{\ua}n^{\ua}\sigma^{2}
+X^{\ua}_{0}+\hat{X}^{\ua}(\tau,\sigma),\non
X^{\oa}&=&2b\delta^{\oa}_{10}\sigma^{1}+\hat{X}^{\oa}(\tau,\sigma),\non
X^{-}&=&-\frac{R}{m}H\tau+2\pi Rn\sigma^{2}
+\hat{X}^{-}(\tau,\sigma), \non
\theta^{(-)}&=&\theta^{(-)}_{0}+\hat{\theta}^{(-)}(\tau,\sigma),
\quad \theta^{(+)}=\hat{\theta}^{(+)}(\tau,\sigma),
\ea
where the hat  stands for the oscillating part.
In this notation, constraints are rewritten as
\ba
0&\approx&\varphi(\sigma)=\nabla^{a}\left(\frac{\hat{P}^{a}}{\sqrt{w}}\right)
+\{\hat{X}^{a},\frac{\hat{P}^{a}}{\sqrt{w}}\}
{}-\frac{i}{2}\{\hat{\theta}^{T},\hat{\theta}\},\non
0&\approx&\varphi_{2}=2\pi(nm+n^{\ua}m^{\ua})+\int d^{2}\sigma
(\hat{P}^{a}\partial_{2}\hat{X}^{a}+\frac{i}{2}\sqrt{w}\hat{\theta}^{T}
\partial_{2}\hat{\theta}),
\ea
where $\nabla^{\ua}\equiv-\pi R^{\ua}n^{\ua}\partial_{1}$ and
$\nabla^{\oa}\equiv b\delta^{\oa}_{10}\partial_{2}$.
We can also calculate membrane charges
\ba
Z^{\oa\ob}&=&Z^{\ua\ub}=Z^{\ua}=0,\non
Z^{\oa\ub}&=&-(2\pi bR^{\ub}n^{\ub})\delta^{\oa}_{10},\non
Z^{\oa}&=&-(2\pi bRn)\delta^{\oa}_{10}.
\ea

Now we can identify the BPS configurations.
Let us start with the configuration which preserves 1/4 SUSY.
Such a BPS configuration should make the matrix ${\bf m}$ 
(eq.(\ref{eq:determinant})) vanish.
Because the last term in eq.(\ref{eq:determinant}) always vanishes
in the cylindrical membrane, we have the following BPS conditions
\ba
0&=&{\cal M}^{2}-Z^{\oa\ub}Z^{\oa\ub}\non
&=&\int d^{2}\sigma\sqrt{w}\left[(\hat{P}^{a}/\sqrt{w})^{2}+
\frac{1}{2}(\{X^{a},X^{b}\}+Z^{ab})^{2}- i \hat{\theta}^{T}
\gamma^{a}\{X^{a},\hat{\theta}\}\right],\non
0&=&Z^{\oa}P^{+}_{0}+Z^{\oa\uc}P^{\uc}_{0}
=-2\pi b\delta^{\oa}_{10}(nm+n^{\uc}m^{\uc})\non
&\approx&b\delta^{\oa}_{10}\int d^{2}\sigma(
\hat{P}^{a}\partial_{2}\hat{X}^{a}+\frac{i}{2}\sqrt{w}
\hat{\theta}^{T}\partial_{2}\hat{\theta}).
\ea
As a general solution to these conditions,
we find the BPS configuration with 1/4 SUSY:
\ba
X^{\ua}&=&\frac{Rm^{\ua}}{R^{\ua}m}\tau+2\pi R^{\ua}n^{\ua}\sigma^{2},
\quad X^{\oa}=2b\delta^{\oa}_{10}\sigma^{1},\non
X^{-}&=&-\frac{R}{m}H\tau+2\pi Rn\sigma^{2}
\quad(\mbox{with }mn+n^{\ua}m^{\ua}=0),\non
\theta^{(-)}&=&\theta^{(-)}_{0},\quad \theta^{(+)}=0.
\ea
It represents a (2+1)-dimensional hyperplane-like membrane
which stretches between the two five-branes. It
should be closely related to the
``intersecting-brane'' configurations\cite{r:Tsey}.
A matrix version of this configuration corresponds to the
``open membrane in M(atrix) theory''\cite{r:Li}.

Next we consider the configuration with $1/8$ SUSY.
In such a configuration the rank of the $16\times 16$ matrix
${\bf m}$ becomes 4.
The BPS bound is given by
\be
{\cal M}^{2}-Z^{\oa\ub}Z^{\oa\ub}\mp2(Z^{10}P^{+}_{0}+
Z^{10\uc}P^{\uc}_{0})=0.
\ee
The analysis is almost parallel to that of
BPS states with 1/4 SUSY for the closed supermembrane\cite{r:EMM}.
We find, in the case $\hat{\theta}^{(\pm)}=0$,
the following BPS conditions
\ba
\hat{P}^{10}&=&0,\non
\frac{\hat{P}^{i}}{\sqrt{w}}&=&
\pm(\{X^{10},X^{i}\}+Z^{10i}),\non
0&=&\{X^{i},X^{j}\}+Z^{ij},\label{eq:1/8BPS}
\ea
where $i,j=2,3,\ldots,9$. The generator of unbroken SUSY is
given by
\be
\tilde{Q}^{(\mp)}\equiv{\cal P}^{(+)}\frac{1\mp\gamma_{10}}{2}
\int d^{2}\sigma\left[\hat{P}^{a}\gamma_{a}+
\frac{\sqrt{w}}{2}(\{X^{a},X^{b}\}+Z^{ab})\gamma_{ab}\right]\hat{\theta}.
\ee
We can see that eq.(\ref{eq:1/8BPS}) is equivalent to the condition
$(\tilde{Q}^{(\mp)},\hat{\theta})_{DB}=0$.\footnote{
One might claim that the latter only imposes
$\{X^{\ua},X^{\ub}\}=\frac{1}{2}\epsilon^{\ua\ub\uc\ud}
\{X^{\uc},X^{\ud}\}$ instead of $\{X^{\ua},X^{\ub}\}=0$
when the target space is $T^d$ with $d\geq 4$.
However, these two equations turn out to be equivalent
by virtue of the identity
$$
\int d^{2}\sigma\sqrt{w}\epsilon^{\ua\ub\uc\ud}\{X^{\ua},X^{\ub}\}
\{X^{\uc},X^{\ud}\}=0.
$$}

As in ref.\cite{r:EMM} we can provide
an example of the configurations with 1/8 SUSY
with nonvanishing $\hat{\theta}$.\footnote{As a matter of fact we can show
that this example yields an almost general solution to the BPS conditions
(\ref{eq:1/8BPS}).  For a detailed analysis, we refer the reader
to Appendix E.}
We consider the following
``dimensional reduction'' of the membrane world-volume
\ba
X^{\bar{a}}&=&2b\delta^{\bar{a}}_{10}\sigma^{1},\non
X^{\ua}&=&\frac{Rm^{\ua}}{R^{\ua}m}\tau+2\pi R^{\ua}n^{\ua}\sigma^{2}
+\hat{X}^{\ua}(\tau,\sigma^{2}),\non
X^{-}&=&-\frac{R}{m}H\tau+2\pi
Rn\sigma^{2}+\hat{X}^{-}(\tau,\sigma^{2}),
\non
\theta^{(+)}&=&0,\non
\theta^{(-)}&=&\theta^{(-)}_{0}+\hat{\theta}^{(-)}(\tau,\sigma^{2}).
\ea
The constraints and
the BPS conditions are reduced to the following form:
\ba
0&\approx&\varphi_{2}=2\pi (nm+n^{\ua}m^{\ua})
  +\oint d\sigma^{2}\left( \frac{\hat{P}^{\ua}}{\sqrt{w}}\partial_{2}
   \hat{X}^{\ua}+\frac{i}{2}\hat{\theta}^{(-)T}\partial_{2}
   \hat{\theta}^{(-)}\right), \non
\partial_{\tau}\hat{X}^{\ua}&=&\pm\frac{Rb}{m}\partial_{2}
   \hat{X}^{\ua} , \non
\hat{\theta}^{(-)}&=&\mp\gamma_{10}\hat{\theta}^{(-)}.
\ea
This configuration
represents an interval times a closed string which is composed only
of the right-(left-)moving modes.
In the limit $b\rightarrow 0$ it yields a tensionless string
which is static on the five-brane world volume.

\section{Discussions}

In this paper we have investigated open supermembranes
in the 11D rigid superspace. 
We have seen that kappa-symmetry and invariance under a
fraction of 11D SUSY  specify the Dirichlet boundary
conditions. The conditions for the fermion
seem to enforce the ``self-duality''
of the three-form field strength on the five-brane
world volume.
In retrospect,  kappa-symmetry of the closed supermembrane
in a curved background required the background be a solution
of 11D supergravity\cite{r:BST}.
In this sense kappa-symmetry of the supermembrane theory
in the covariant formalism plays a role similar to that of the
conformal invariance in superstring theory.
This reasoning leads us to the conjecture that
kappa-symmetry of open supermembranes in a curved background
yields the field equations for the (collective modes of) M-theory
five-branes. It would be interesting to pursue this possibility.
\footnote{Quite recently Chu and Sezgin demonstrated that
our conjecture is indeed true \cite{r:ChuSez}.}

We have also shown that the light-cone gauge formulation is
regularized by a (0+1)-dimensional SO(N) supersymmetric gauge theory.
It is known that
the matrix regularization of the closed supermembrane is
closely related to the matrix formulation of
M-theory \cite{r:BFSS}. Because our SO(N$\rightarrow\infty$) theory
describes an open supermembrane and a five-brane which are also
essential to the description of M-theory,
it is conceivable that the true M(atrix) theory incorporates
naturally the SO(N$\rightarrow\infty$) theory
in a certain sense.

We should remark that our analysis is classical
and thus we do not consider the effect of
anomalies. Because the boundary of the open membrane
is two dimensional and because the fields on the five-brane world
volume is chiral, anomalies are expected to arise\cite{r:Wit5}\cite{r:BM}.
It is an important task to examine what modification is required
if anomalies are taken into account.\footnote{
A related topic is the M(atrix) theory compactified on
$T^{5}/{\bf Z}_{2}$ which is described by a
USp(2N) supersymmetric gauge theory\cite{r:FS}.
It might be worthwhile constructing an anomaly-free theory
by combining this USp(2N) model with our
SO(N) model.}
\footnote{Recently Brax and Mourad have analyzed in detail
the issue of anomalies in the theory of open supermembranes
which end on the five-branes\cite{r:BM2}.}
In particular, it would be of interest to see whether the
$p=1$ case survives through anomalies.

While we have concentrated on the $p=5$ case, our result
may be extended to the $p=9$ case. This case is interesting
because it is related to the Matrix theory of heterotic strings
\cite{r:DF}\cite{r:KR}.
{}Finally we will briefly discuss their relation.
Boundary conditions for the open supermembrane
which ends on the 9-brane are given by
\ba
\mbox{Dirichlet}&:& \omega,\quad X^{10},
\quad\frac{1-\Gamma_{(9)}}{2}\theta;\non
\mbox{Neumann}&:& X^{\ua}\quad(\ua=2,3,\ldots,9),
\quad \frac{1+\Gamma_{(9)}}{2}\theta,
\ea
After performing the matrix regularization this agrees with
the matter contents of \cite{r:DF}\cite{r:KR} because  Dirichlet and
Neumann modes are respectively approximated by $N\times N$ antisymmetric
and symmetric matrices. This gives a support for the idea
of the heterotic matrix theory from a different point of view.
It would therefore be intriguing to further investigate this model
{}from the membrane side.

\vskip 5mm
\noindent{\bf Acknowledgements:} \hskip 3mm
We would like to thank
M. Ninomiya, M. Anazawa, A. Ishikawa, K. Sugiyama and S. Watamura
for invaluable discussions, comments, and encouragement.
We are also grateful to the referees of Physical Review D
who provided us useful advice for improving the manuscript.


\appendix

\section{Convention and useful formulae}

\subsection{SO(10,1) Clifford algebra}

Eleven dimensional gamma matrices $\Gamma^{\mu}$
($\mu=0,1,\ldots,10$) satisfy the SO(10,1) Clifford algebra
\be
\Gamma^{\mu}\Gamma^{\nu}=\eta^{\mu\nu}I_{32}+\Gamma^{\mu\nu},
\ee
where $(\eta^{\mu\nu})=\mbox{diag}(-,+,\ldots,+)$ denotes the
eleven dimensional Minkowski metric and we use the notation
$$
\Gamma^{\mu_{1}\cdots\mu_{n}}\equiv\Gamma^{[\mu_{1}}\cdots\Gamma^{\mu_{n}]}.
$$
We should remark that $\Gamma^{10}$ is identified with the ten
dimensional chiral matrix $\Gamma_{11}=\Gamma_{01\cdots 9}$.
If we define the charge conjugation matrix ${\cal C}$ by
\be
{\cal C}^{-1}(\Gamma^{\mu})^{T}{\cal C}=-\Gamma^{\mu},
\ee
the gamma matrices have the following properties
\ba
&&{\cal C}^{-1}(\Gamma^{\mu_{1}\cdots\mu_{n}})^{T}{\cal C}
=(-)^{\frac{n(n+1)}{2}}\Gamma^{\mu_{1}\cdots\mu_{n}}, \non
&&({\cal C}\Gamma_{\mu\nu})_{(\alpha\beta}
({\cal C}\Gamma^{\nu})_{\gamma\delta)}=0.
\ea
{}From the first equation it follows
\be
{\cal C}^{-1}(\Gamma_{(p)})^{T}{\cal C}=(-)^{p}(\Gamma_{(p)})^{-1},
\ee
where $\Gamma_{(p)}\equiv\Gamma_{01\cdots p}$.

In practical calculation it is frequently convenient
to use the Majorana representation in which the spinor
is real and
\be
\Gamma^{0}={\cal C}=\left(\begin{array}{cc}0&I_{16}\\-I_{16}&0
\end{array}\right),\quad
\Gamma^{1}=\left(\begin{array}{cc}0&-I_{16}\\-I_{16}&0
\end{array}\right), \quad
\Gamma^{a}=\left(\begin{array}{cc}\gamma^{a}&0\\0&-\gamma^{a}
\end{array}\right),
\ee
where $\gamma^{a}$ ($a=2,3,\ldots,10$) are SO(9) gamma matrices
in the real representation in which $\gamma^{a}$ are real and symmetric.

\subsection{SO(5,1)$\subset$SO(10,1) Clifford algebra}

Among the eleven dimensional gamma matrices,
$\Gamma^{\umu}$ ($\umu=0,1,\ldots,5$) form an SO(5,1)
Clifford algebra. The 6D chirality is defined by the matrix
\be
\Gamma_{(5)}=\Gamma_{01\cdots 5}=\frac{\epsilon^{\umu\unu\urho\usig
\ukap\ulam}}{6!}\Gamma_{\umu\unu\urho\usig\ukap\ulam},
\label{eq:gamma5}
\ee
which satisfies
\be
(\Gamma_{(5)})^{2}=I_{32},
\quad \Gamma_{(5)}\Gamma^{\umu}+\Gamma^{\umu}\Gamma_{(5)}=0.
\ee
{}From the definition (\ref{eq:gamma5})
we find the `duality' relations
which are useful in the analysis of open supermembranes
\ba
\Gamma_{\umu\unu\urho\usig\ukap\ulam}&=&
{}-\epsilon_{\umu\unu\urho\usig\ukap\ulam}\Gamma_{(5)},\non
\Gamma_{\umu\unu\urho\usig\ukap}&=&
{}-\epsilon_{\umu\unu\urho\usig\ukap\ulam}\Gamma_{(5)}\Gamma^{\ulam},\non
\Gamma_{\umu\unu\urho\usig}&=&\frac{1}{2}
\epsilon_{\umu\unu\urho\usig\ukap\ulam}
\Gamma_{(5)}\Gamma^{\ukap\ulam},\non
\Gamma_{\umu\unu\urho}&=&\frac{1}{3!}
\epsilon_{\umu\unu\urho\usig\ukap\ulam}
\Gamma_{(5)}\Gamma^{\usig\ukap\ulam}.
\ea

\subsection{Self-dual three-form $h_{\umu\unu\urho}$}

In the analysis of open supermembranes we frequently deal with
the self-dual three-form on the five-brane world volume
\be
h_{\umu\unu\urho}=\frac{1}{3!}\epsilon_{\umu\unu\urho\usig\ukap\ulam}
h^{\usig\ukap\ulam}.
\ee
If we define the
tensor $k^{\umu}_{\unu}\equiv h^{\umu\urho\usig}h_{\unu\urho\usig}$,
we find the following useful identities
\ba
k^{\umu}_{\umu}&=&h^{\umu\unu\urho}h_{\umu\unu\urho}=0,\non
h^{\umu\unu\ukap}h_{\urho\usig\ukap}
&=&\delta^{[\umu}_{[\urho}k^{\unu]}_{\usig]},\non
k^{\umu}_{\usig}h^{\usig\unu\urho}&=&k^{\unu}_{\usig}
h^{\umu\usig\urho},\non
{h^{\umu[\unu}}_{\ukap}h^{\urho\usig]\ukap}&=&0,\non
k_{\umu}^{\urho}k_{\urho}^{\unu}&=&\frac{1}{6}\delta^{\unu}_{\umu}
(k^{\urho}_{\usig}k^{\usig}_{\urho}),\non
k_{\umu}^{\usig}k_{\unu}^{\ukap}h_{\usig\ukap\urho}&=&
\frac{1}{6}(k_{\ukap}^{\ulam}k_{\ulam}^{\ukap})h_{\umu\unu\urho}.
\ea

\section{Derivation of eq.(28)}

In this appendix we reproduce eq.(\ref{eq:Dirichlet})
{}from eq.(\ref{eq:linearD}). We first note that
(\ref{eq:linearD}) is rewritten as
\be
\theta^{\prime}\equiv
\exp(\frac{1}{3!}h_{\umu\unu\urho}\Gamma^{\umu\unu\urho})\theta
=\Gamma_{(5)}\theta^{\prime}.
\ee
Thus we find
\ba
0&=&\bar{\theta^{\prime}}\Gamma_{\umu\unu}\delta\theta^{\prime}\non
&=&\bar{\theta}\exp\left(\frac{1}{3!}h_{\urho\usig\ukap}
\Gamma^{\urho\usig\ukap}\right)
\Gamma_{\umu\unu}\exp\left(\frac{1}{3!}
h_{\urho\usig\ukap}\Gamma^{\urho\usig\ukap}\right)\delta\theta  \non
&=&\bar{\theta}\Gamma_{\umu\unu}\delta\theta-2h_{\umu\unu\urho}
\bar{\theta}\Gamma^{\urho}(1+\Gamma_{(5)})\delta\theta.
\label{eq:int1}
\ea
Here we have used the equations in Appendix A.3 and
$(h_{\umu\unu\urho}\Gamma^{\umu\unu\urho})^{2}=0$.
To rearrange this equation into the desired form
we have to express $\bar{\theta}\Gamma^{\umu}\Gamma_{(5)}\delta\theta$
in terms of $\bar{\theta}\Gamma^{\umu}\delta\theta$ and of
$\bar{\theta}\Gamma_{\umu\unu}\delta\theta$. For this purpose
we rewrite eq.(\ref{eq:linearD}) as
\be
\Gamma_{(5)}\theta=(1+\frac{1}{3}h_{\umu\unu\urho}\Gamma^{\umu\unu\urho})
\theta.
\ee
Using this equation and equations in Appendix A.3 we find
\be
(1+2k)^{\umu}_{\unu}\Gamma^{\unu}\Gamma_{(5)}\theta=
(1-2k)^{\umu}_{\unu}\Gamma^{\unu}\theta+2h^{\umu\unu\urho}
\Gamma_{\unu\urho}\theta.
\ee
Substituting it into eq.(\ref{eq:int1}) yields
\be
\bar{\theta}\Gamma_{\umu\unu}\delta\theta-k_{\umu}^{\urho}
\bar{\theta}\Gamma_{\urho\unu}\delta\theta-k_{\unu}^{\urho}
\bar{\theta}\Gamma_{\umu\urho}\delta\theta=4h_{\umu\unu\urho}
\bar{\theta}\Gamma^{\urho}\delta\theta.
\ee
An inspection shows that this is equivalent to the equation
\ba
\bar{\theta}\Gamma_{\umu\unu}\delta\theta&=&4(1-\frac{2}{3}k^{\urho}_{\usig}
k_{\urho}^{\usig})^{-1}
(h_{\umu\unu\ukap}+k_{\umu}^{\ulam}h_{\ulam\unu\ukap}
+k_{\unu}^{\ulam}h_{\umu\ulam\ukap})\bar{\theta}\Gamma^{\ukap}\delta\theta
\non
&=&4h_{\umu\unu\usig}{(1-2k)^{-1}}^{\usig}_{\urho}
\bar{\theta}\Gamma^{\urho}\delta\theta,
\ea
which is identical to the last condition of (\ref{eq:Dirichlet})
providing eq.(\ref{eq:Hh}) holds.

\section{Matrix approximation of Dirichlet and Neumann modes}

In this section we show that Dirichlet and Neumann modes
are well approximated by $N\times N$ antisymmetric and symmetric
matrices, respectively.
We recall\cite{r:dWMN}\cite{r:FFZ} that the Fourier modes
on the torus (parametrized by $(\sigma^{1},\sigma^{2})\in
[0,1)^{2}$) is approximated as
\be
Y_{A}(\sigma)\equiv e^{2\pi i(A_{1}\sigma^{1}+A_{2}\sigma^{2})}
\stackrel{N\rightarrow\infty}{\longleftarrow}
T_{A}\equiv\frac{1}{\sqrt{N}}e^{-\frac{\pi
i}{N}A_{1}A_{2}}V^{A_{1}}U^{A_{2}},
\ee
where 't Hooft's twist matrices $U$ and $V$ have the following
properties\cite{r:thooft}
\be
U^{N}=V^{N}=1,\quad VU=e^{\frac{2\pi}{N}i}UV, \quad U^{\dagger}=U^{-1}
,\quad V^{\dagger}=V^{-1}.
\ee
The phase factor of $T_{A}$ is chosen so that we have
\be
(T_{A})^{\dagger}=T_{-A}.
\ee
We take the following representation for $U,V$
\be
U=\left(\begin{array}{ccccc}
1& & & & 0\\
 &e^{\frac{2\pi}{N}i}& & & \\
 & &\ddots & & \\
 & & &\ddots& \\
0& & & &e^{\frac{2\pi i}{N}(N-1)} 
\end{array}\right),\qquad
V=\left(\begin{array}{ccccc}
0&1& & &0\\
 &0&1& & \\
 & &\ddots&\ddots& \\
 & & &\ddots&1\\
1& & & &0
\end{array}\right),\label{eq:twist}
\ee
with the properties $U^{T}=U$ and $V^{T}=V^{-1}$. 
Then we find the important relation
\be\label{e:partialConjugate}
T_{(-A_{1},A_{2})}=(T_{(A_{1},A_{2})})^{T}.
\ee
This enables us to confirm that the following correspondence
holds
\ba
Y^{(D)}_{A}=\frac{-i}{\sqrt{2}}(Y_{(A_{1},A_{2})}-Y_{(-A_{1},A_{2})})
&\stackrel{N\rightarrow\infty}{\longleftarrow}&
T^{(D)}_{A}\equiv\frac{-i}{\sqrt{2}}(T_{A}-(T_{A})^{T}),\non
Y^{(N)}_{A}=\frac{1}{\sqrt{2}}(Y_{(A_{1},A_{2})}+Y_{(-A_{1},A_{2})})
&\stackrel{N\rightarrow\infty}{\longleftarrow}&
T^{(N)}_{A}\equiv\frac{1}{\sqrt{2}}(T_{A}+(T_{A})^{T}).
\ea
Because $T^{(D)}_{A}$ ($T^{(N)}_{A}$) are manifestly
antisymmetric (symmetric), this implies the correspondence
(\ref{eq:correspondence}).

We note that, as far as the representation of SO(N) is concerned,
the correspondence(\ref{eq:correspondence}) is
independent of the choice of twist matrices $(U,V)$,
because any twist matrices are unitary equivalent
to those in eq.(\ref{eq:twist}).
{}For example one may choose $U$ instead of $V$ to define
$e^{2\pi i \sigma_1}$.  In this choice, (\ref{e:partialConjugate})
is replaced by 
\be\label{e:partialConjugate2}
T_{(-A_{1},A_{2})}=(T_{(A_{1},A_{2})})^{*}.
\ee
In this convention, the generators associated with
the Dirichlet modes $\sin(2\pi i n \sigma_1)\sin(2\pi i m\sigma_2)$ 
are anti-symmetric
but those associated with another type
$\sin(2\pi i n \sigma_1) \cos(2\pi i m\sigma_2)$ become symmetric.
In this sense, our claim that the Dirichlet mode is described by
the antisymmetric matrix depends on our specific choice of
the basis.  However, since these different choices are
equivalent under a unitary transformation, 
the underlying algebraic structure remains the same.

\section{Matrix approximation of surface integration}

In this section we examine whether the integration
$\int_{\Sigma^{(2)}}d^{2}\sigma$ possesses the property
of the trace in the restricted situation (\ref{eq:DDNN}).
We first note that, in the situation we consider,
the integration on the cylinder equals to that on the torus, namely
$$
\int^{1/2}_{0}d\sigma^{1}\int^{1}_{0}d\sigma^{2}2F(\sigma)
=\int^{1}_{0}d\sigma^{1}\int^{1}_{0}d\sigma^{2}F(\sigma),
$$
where $F(\sigma^{1},\sigma^{2})=F(1-\sigma^{1},\sigma^{2})$
is a periodic function on the torus.
Since the integration on the torus is well-approximated by
the trace of $N\times N$ matrices, the same should be true for
that on the cylinder as long as we consider the structure (\ref{eq:DDNN}).
{}For completeness  we  show that the cyclic identity
\be
\int_{\Sigma^{(2)}}d^{2}\sigma\sqrt{w}A\{B,C\}=
\int_{\Sigma^{(2)}}d^{2}\sigma\sqrt{w}B\{C,A\}
\ee
holds in the following two cases.

(i)\underline{$A,B,C=$Dirichlet.}
\ba
L.H.S.&=&\int d^{2}\sigma A\epsilon^{rs}\partial_{r}B\partial_{s}C\non
&=&-\int d^{2}\sigma B\epsilon^{rs}\partial_{r}A\partial_{s}C
+\int d\sigma^{2}(AB\partial_{2}C)|^{\sigma^{1}=1/2}_{\sigma^{1}=0}\non
&=&R.H.S..
\ea

(ii)\underline{ $A=$Dirichlet; $B,C=$Neumann.}
\ba
L.H.S.&=&\int d^{2}\sigma\sqrt{w}B\{C,A\}+
\int d\sigma^{2}(AB\partial_{2}C)|^{\sigma^{1}=1/2}_{\sigma^{1}=0}\non
&=&R.H.S..
\ea

Thus we can approximate the integration by the trace
in the situation which appears in the light-cone gauge
formulation of the open supermembrane.

\section{General solutions of BPS equations}

Let us start from the situation 
in ref.\cite{r:EMM} where we have investigated the BPS conditions
for the closed toroidal supermembrane in the
target space which is toroidally compactified, 
i.e., $X^{a}\sim X^{a}+2\pi R^{a}$ $(a=1,\ldots,9)$ and
$X^{-}\sim X^{-}+2\pi R$. In general the embedding
functions and their conjugate momenta are expanded as
\ba
X^- & = & -Ht + 2\pi Rn_{r}\sigma^{r}
+\hat{X}^-(\sigma^{1},\sigma^{2},t)\nonumber \\
X^a & = &\frac{m^a}{R^a} t
+2\pi R^a n^a_r\sigma^r+\hat{X^a}(\sigma^{1},\sigma^{2},t),\nonumber \\
P^a & \equiv & \partial_{t}X^{a}=
\frac{m^a}{R^a} + \hat{P^a} (\sigma^{1},\sigma^{2},t),\nonumber \\
\ea
where we have used the rescaled time $t\equiv\frac{R}{m}\tau$ and
the symbols with hat denote the contributions from the oscillating
modes on $\Sigma^{(2)}\approx T^{2}$.

We have also introduced a nine-dimensional
orthonormal basis $(e^{(9)}_{a},e^{(i)}_{a})$ ($i=1,\ldots,8$)
such that $P^{+}_{0}z^{a}-P^{c}_{0}z^{ca}\propto e^{(9)a}$,
where $z^{a}$ and $z^{ca}$ are longitudinal and transverse
membrane charges, respectively.
We denote the components of a nine-dimensional vector $V^{a}$ as
$$
V^{9}=e^{(9)}_{a}V^{a},\quad V^{i}=e^{(i)}_{a}V^{a}.
$$
For simplicity we set $\hat{\theta}=0$. Extension to the
case of nonzero $\hat{\theta}$ can be carried out if we use
the prescription explained in sec.5 of ref.\cite{r:EMM}. 

In this setup we have shown that the configurations with 1/4 SUSY must
satisfy the BPS conditions
\ba
\hat{P}^9  & = & 0, \label{e:BPS1}\\
\hat{P}^i & = & \pm \left( \left\{ X^9, X^i \right\}
+z^{9i}\right),\label{e:BPS2}\\
0 & = & \left\{ X^i, X^j\right\}
+ z^{ij},\label{e:BPS3}
\ea
as well as the constraints
\ba
0&=&\varphi(\sigma)=\nabla^{a}\hat{P}^{a}+\{\hat{X}^{a},\hat{P}^{a}\},
\label{eq:con1}\\
0&=&\varphi_{r}=mn_{r}+m^{a}n^{a}_{r}+\frac{1}{2\pi}\int d^{2}\sigma
\hat{P}^{a}\partial_{r}\hat{X}^{a},
\label{eq:con2}\\
0&=&P^{+}_{0}z^{i}-P^{c}_{0}z^{ci},\label{eq:redundant}
\ea
where $\nabla^{a}\equiv2\pi R ^{a}(n^{a}_{1}\partial_{2}-
n^{a}_{2}\partial_{1})$.

In the following we
investigate what the general solutions of eqs.(\ref{e:BPS1}-
\ref{eq:redundant}) would look like.
We note that, among these equations,
the consistency condition (\ref{eq:redundant})
is redundant because it automatically holds
as a consequence of eqs.(\ref{e:BPS1}),(\ref{e:BPS3}),(\ref{eq:con1}) and
(\ref{eq:con2}):
\ba
P^{+}_{0}z^{i}-P^{c}_{0}z^{ci}&=&\int d^{2}\sigma\hat{P}^{c}\nabla^{i}
\hat{X}^{c}\nonumber \\
&=&\int d^{2}\sigma\hat{P}^{c}(\nabla^{i}\hat{X}^{c}
-\nabla^{c}\hat{X}^{i}+\{\hat{X}^{i},\hat{X}^{c}\})\nonumber \\
&=&\int d^{2}\sigma\hat{P}^{j}(\{X^{i},X^{j}\}+z^{ij})\nonumber \\
&=&0.
\ea 

It is trivial to solve the first BPS condition
(\ref{e:BPS1}). The general solution is,
\be
X^9 = P^{9}_{0}t + 2\pi\tilde{R}^{9}_{r}\sigma^{r} +
\xi(\sigma^1,\sigma^2),
\ee
where $\tilde{R}^{9}_{r}\equiv\sum_{a=1}^{9}e^{(9)}_{a}R^{a}n^{a}_{r}$.
Assuming that $\xi$ is sufficiently small, a local APD gauge transformation
further reduces this to the following form
\be
X^9 = P^{9}_{0}t + 2\pi\tilde{R}^{9}_{r}\sigma^{r} +
\xi(\tilde{R}^{9}_{r}\sigma^{r}).\label{eq:X9}
\ee

In general, nonlinear partial differential equation
(\ref{e:BPS3}) is hard to solve.  Here we consider
physically interesting situation
where there exist nonvanishing windings
$\tilde{R}^{i}_{r}\equiv\sum^{9}_{a=1}
e^{(i)}_{a}R^{a}n^{a}_{r}$ and the oscillating modes
$\hat{X}^{i}$ are infinitesimally small.
In this case the equation reduces to the linear partial differential
equation,
\be
\nabla^i \hat{X}^j - \nabla^j \hat{X}^i = 0,
\quad
\nabla^i \equiv e^{(i)}_{a}\nabla^{a}=2\pi(
\tilde{R}^{i}_{1}\partial_{2}-\tilde{R}^{i}_{2}\partial_{1}).
\label{e:BPS4}
\ee

The linearlized version can be straightfowardly solved.
The general solution to the equation for one pair (say $(i,j)$)
is given as follows,
\ba
\hat{X}^i &=& \nabla^i \epsilon_{ij}(\sigma^{1},\sigma^{2},t)
+ \eta^i(\sigma^{1},\sigma^{2},t),\nonumber \\
\hat{X}^j &=& \nabla^j \epsilon_{ij}(\sigma^{1},\sigma^{2},t)
+ \eta^j(\sigma^{1},\sigma^{2},t),
\ea
where $\eta^i$, $\eta^j$ satisfy
\be
\nabla^j \eta^i= \nabla^i \eta^j = 0.
\ee
The analysis has to be made separately for the two cases,
$z^{ij}=0$ and $z^{ij}\neq 0$.  But the final answer can
be written in the same way.

Suppose that we need to consider $T^d$ with $d>2$, we have to
study the consistency condition among the solutions for every
pair.
It seems that we need to make classification of the solution
into two cases.

\noindent (a) Every $z^{ij}$ vanishes.  In this case, we can factorize
$\tilde{R}^{i}_{r}$ as $\tilde{R}^{i}_{r}=k^{i}\tilde{R}_{r}$ and
we can take
$\eta^i\neq 0$.  This $\eta$ is the string excitation
in the double dimensional reduction.  $\epsilon$s are constrained
to be the same for every pair, namely $\epsilon^{ij} = \epsilon$.
Thus we find
\be
X^{i}=P^{i}_{0}t+2\pi k^{i}\tilde{R}_{r}\sigma^{r}
+\nabla^{i}\epsilon(\sigma^{1},\sigma^{2},t)
+\eta^{i}(\tilde{R}_{r}\sigma^{r},t).\label{eq:Xi1}
\ee

\noindent (b) There are some nonvanishing $z^{ij}$.  In this case,
we need to put every $\eta$ to vanish.  As before every 
$\epsilon$ are the same $\epsilon^{ij} = \epsilon$. The result is
\be
X^{i}=P^{i}_{0}t+2\pi\tilde{R}^{i}_{r}\sigma^{r}
+\nabla^{i}\epsilon(\sigma^{1},\sigma^{2},t).\label{eq:Xi2}
\ee

Next we solve the linearized version of the
constraint (\ref{eq:con1}), namely, $\nabla^{i}\hat{P}^{i}=0$.
By substisuting (\ref{eq:Xi1}) or (\ref{eq:Xi2}) into
this equation, we find that $\epsilon$ is time-independent,
apart from the part which can be absorbed into $\eta^{i}$.

The equation
(\ref{e:BPS2}) defines the time evolution of remaining unknown
quantities $\epsilon$ and $\eta$. It is easier to study the case
(b). The result of substituting (\ref{eq:X9}) and (\ref{eq:Xi2})
into (\ref{e:BPS2}) is given by $0=\nabla^{i}[\nabla^{9}
\epsilon(\sigma^{1},\sigma^{2})-\xi(\tilde{R}^{9}_{r}\sigma^{r})].$
{}From this we can easily see
\be
\nabla^{9}\epsilon(\sigma^{1},\sigma^{2})-
\xi(\tilde{R}^{9}_{r}\sigma^{r})=0.
\ee
In order for this to hold we have to set
\be
\xi=0,\qquad  \epsilon=\epsilon(\tilde{R}^{9}_{r}\sigma^{r}).
\ee
We can absorb this $\epsilon$ by an infinitesimal APD.
The final result is
\be
X^{9}=P^{9}_{0}t+2\pi\tilde{R}^{9}_{r}\sigma^{r},
\quad
X^{i}=P^{i}_{0}t+2\pi\tilde{R}^{i}_{r}\sigma^{r}.
\ee
This is nothing but the BPS configuration with 1/2 SUSY.

Let us now examine the case (a). By substituting
(\ref{eq:X9}) and (\ref{eq:Xi1}) into (\ref{e:BPS2}), we find
\be
(\partial_{t}\mp\nabla^{9})\eta^{i}(\tilde{R}_{r}
\sigma^{r},t)=
\pm k^{i}\nabla(\nabla^{9}\epsilon(\sigma^{1},\sigma^{2})
-\xi(\tilde{R}^{9}_{r}\sigma^{r})),
\ee 
where $\nabla\equiv2\pi(\tilde{R}_{1}\partial_{2}
-\tilde{R}_{2}\partial_{1})$.
This implies the following equations
\ba
(\frac{m}{R}\partial_{t}\mp\nabla^{9})\eta^{i}&=&0,
\label{eq:string} \\
\epsilon(\sigma^{1},\sigma^{2})&=&\epsilon^{(1)}(\tilde{R}_{r}\sigma^{r})
+\epsilon^{(2)}(\tilde{R}^{9}_{r}\sigma^{r}),\\
\xi(\tilde{R}^{9}_{r}\sigma^{r})&=&0.
\ea
Plugging these back into eqs.(\ref{eq:X9})and (\ref{eq:Xi1})
and performing an appropriate APD transformation we find
\ba
X^{9}&=&P^{9}_{0}t+2\pi\tilde{R}^{9}_{r}\sigma^{r},
\nonumber \\
X^{i}&=&P^{i}_{0}t+2\pi k^{i}\tilde{R}_{r}\sigma^{r}
+\eta^{i}(\tilde{R}_{r}\sigma^{r}).\label{eq:caseIa}
\ea

We note that, if the ratio $\tilde{R}_{1}/\tilde{R}_{2}$
is irrational, $\eta^{i}$ in eq.(\ref{eq:caseIa}) become
constant and the solution
reduces to that with 1/2 SUSY. Thus, in order to obtain
the nontrivial BPS configuration with 1/4 SUSY,
we have to take this ratio to be rational.
In this case, by using an appropriate $SL(2,{\bf Z})$
transformation
\be
\left(\begin{array}{c}
\sigma_1\\ \sigma_2 \end{array}\right)
= 
\left(\begin{array}{cc}
a & b \\
c & d
\end{array}\right)
\left(\begin{array}{c}
\sigma'_1\\ \sigma'_2 \end{array}\right)\; ,\quad
\left(\begin{array}{cc}
a & b \\
c & d
\end{array}\right)\in\;SL(2,{\bf Z}),
\ee
we can rewrite eq.(\ref{eq:caseIa}) in the form
\ba
X^{9}&=&P^{9}_{0}t+2\pi\tilde{R}^{\prime 9}_{r}\sigma^{r},
\nonumber \\
X^{i}&=&P^{i}_{0}t+2\pi k^{i}\tilde{R}^{\prime}\sigma^{1}
+\hat{X}^{i}(\sigma^{1},t),
\ea
where $\hat{X}^{i}$ satisfy the equation
\be
\partial_{t}\hat{X}^{i}=\mp2\pi\tilde{R}^{9}_{2}\partial_{1}
\hat{X}^{i}.
\ee
This final result is nothing but the extended version
of the ``stringy'' configuration
given in ref.\cite{r:EMM}, with the only extension
being the winding of the $\sigma^{1}$-cycle in the $X^{9}$-direction.

The extension of the above results to the open supermembrane
is straightforward.
Note that we have $2\pi(\tilde{R}^{9}_{1},\tilde{R}^{9}_{2})=(b,0)$
and $\sigma^{1}$ in the above analysis corresponds to
$2\sigma^{1}$ in subsec. IV A.
We only have to inspect what constraints are imposed  by
the bondary conditions.
It turns out that the resulting configuration is given by
eqs.(83) and (84).



\end{document}